\documentclass[10pt,journal,compsoc]{IEEEtran}
\ifCLASSINFOpdf
\else
\fi

\usepackage{amsmath,amsfonts}
\usepackage{algorithmic}
\usepackage{algorithm}
\usepackage{array}
\usepackage{textcomp}
\usepackage{stfloats}
\usepackage{url}
\usepackage{verbatim}
\usepackage{graphicx}
\usepackage{cite}

\usepackage{subfigure}
\usepackage[dvipsnames]{xcolor} 
\usepackage{multirow}
\usepackage{tabularx}
\usepackage{booktabs}

\DeclareMathOperator*{\argmin}{\arg\min}
\DeclareMathOperator*{\argmax}{\arg\max}

\newcommand{\rob}{\mathrm{rob}}
\newcommand{\nat}{\mathrm{nat}}

\newcommand{\cB}{\mathcal{B}}
\newcommand{\cF}{\mathcal{F}}
\newcommand{\cX}{\mathcal{X}}
\newcommand{\cY}{\mathcal{Y}}
\newcommand{\bx}{{x}}
\newcommand{\bxprime}{{x}'}
\newcommand{\bxtidle}{\tilde{{x}}}

\newcommand{\A}{\mathcal{A}}
\newcommand{\R}{\mathbb{R}}
\newcommand{\epsball}{\mathcal{B}_\epsilon}
\newcommand{\epsballzero}{\mathcal{B}_{\epsilon_{0}}}
\newcommand{\xadv}{\tilde{{x}}}
\newcommand{\xtar}{x_{\mathrm{tar}}}
\newcommand{\tar}{\mathrm{tar}}
\newcommand{\adv}{\mathrm{adv}}
\newcommand{\ML}{\mathrm{ML}}
\newcommand{\poi}{\mathrm{poi}}

\begin{document}
%
\title{Assessing Vulnerabilities of Adversarial Learning Algorithm through Poisoning Attacks}
%
%
%
%

\author{
	Jingfeng Zhang*,~
        Bo Song*,~
        Bo Han,~
        Lei Liu,~
        Gang Niu,~
        Masashi Sugiyama
\IEEEcompsocitemizethanks{
\IEEEcompsocthanksitem *The first two authors made the equal contributions.
\IEEEcompsocthanksitem Corresponding author is Jingfeng Zhang. Email: jingfeng.zhang9660@gmail.com \\ 
Jingfeng Zhang, Bo Han, Gang Niu, and Masashi Sugiyama are with RIKEN AIP, Japan. \\
Bo Song and Lei Liu are with Shandong University, Jinan, China.\\
Bo Han is also with Hong Kong Baptist University, Hong Kong, China. 
Masashi Sugiyama is also with the University of Tokyo, Japan. 
}
\thanks{Preprint is currently under review.}
}
\IEEEtitleabstractindextext{%
\begin{abstract}
\textit{Adversarial training} (AT) is a robust learning algorithm that can defend against adversarial attacks in the inference phase and mitigate the side effects of corrupted data in the training phase. As such, it has become an indispensable component of many artificial intelligence (AI) systems. 
However, in high-stake AI applications, it is crucial to understand AT's vulnerabilities to ensure reliable deployment.
In this paper, we investigate AT's susceptibility to poisoning attacks, a type of malicious attack that manipulates training data to compromise the performance of the trained model.
Previous work has focused on poisoning attacks against \textit{standard training}, but little research has been done on their effectiveness against AT. 
To fill this gap, we design and test effective poisoning attacks against AT. 
Specifically, we investigate and design clean-label poisoning attacks, allowing attackers to imperceptibly modify a small fraction of training data to control the algorithm's behavior on a specific target data point. Additionally, we propose the clean-label untargeted attack, enabling attackers can attach tiny \textit{stickers} on training data to degrade the algorithm's performance on all test data, where the stickers could serve as a signal against unauthorized data collection. Our experiments demonstrate that AT can still be poisoned, highlighting the need for caution when using vanilla AT algorithms in security-related applications. The code is at https://github.com/zjfheart/Poison-adv-training.git.
\vspace*{-2mm}
\end{abstract}

\begin{IEEEkeywords}
Poisoning attacks, vulnerabilities of the AI algorithms, adversarial learning algorithm
\vspace*{-2mm}
\end{IEEEkeywords}}

\maketitle

\IEEEdisplaynontitleabstractindextext

%
\IEEEpeerreviewmaketitle

\IEEEraisesectionheading{\section{Introduction}\label{sec:introduction}}

%
%
%
%
\IEEEPARstart{A}rtificial Intelligent (AI) algorithms are increasingly being used in safety-critical systems, such as drones~\cite{chen_drones_manipuations} and healthcare systems~\cite{heath_care_system}, where the robust AI algorithms are critical to enhancing the system's securities. 
{Adversarial training} (AT)~\cite{Goodfellow14_Adversarial_examples} is a widely used robust AI algorithm that trains a model on adversarial data generated within a bounded distance of their natural counterparts~\cite{Madry_adversarial_training,Chaowei_iclr18_deformation}. 
 Previous research on AT has focused on two main objectives: improving the natural accuracy of classification~\cite{zhang2020fat} and making decision boundaries more robust~\cite{yang2020closer,wang2020improving_MART,zhang2021geometryaware,chen_guided_AT}. Achieving these objectives offers two important benefits.

AT has emerged as a promising defense against adversarial attacks~\cite{Athalye_ICML_18_Obfuscated_Gradients} that pose a significant threat to safety-critical AI applications such as medicine and autonomous driving. Attackers can add imperceptible noise to natural data to evade the model's predictions~\cite{Goodfellow14_Adversarial_examples}. Athalye et al. (2018)~\cite{Athalye_ICML_18_Obfuscated_Gradients} showed AT stands out as a promising defense against powerful optimization-based attacks~\cite{dong2018boosting,croce2020reliable,dong_attack_tpami,sriramanan2020guided}.



In addition, AT also offers robustness against corrupted data during the training phase compared to standard training (ST)~\cite{arpit2017closer}. Tao et al. (2021)~\cite{tao_delusive} demonstrated that AT can defend against delusive attacks where attackers imperceptibly poison input features of training data to degrade the model's generalization. Furthermore, Huang et al. (2021)~\cite{huang2021unlearnable} generated unlearnable data that degrade ST's generalization, but Fu et al. (2022)~\cite{fu2022robust} showed that AT is nearly immune to such attacks. Overall, AT mitigates the side effects of corruption in the training set and offers better robustness against adversarial attacks.

\noindent\textbf{Contribution.} 
While AT has been shown to offer robustness benefits in both training and inference phases, few works have actively explored AT's vulnerabilities, which may pose risks in security-related applications.
To address this gap, we propose two types of poison attacks against AT: clean-label targeted attacks~\cite{Shafahi_nips18,huang_matapoison_nips20,geiping_witches_2021} and clean-label untargeted attacks~\cite{zhiqi_privacy,feng_nips19,shan2020protecting,huang2021unlearnable,fowl_adv_strong_poi} against AT. 
These attacks aim to uncover AT's vulnerabilities and increase awareness of potential risks when using AT in AI systems.

\begin{figure*}[tp!]
	\centering
	\subfigure[Clean-label targeted attack]{
		\includegraphics[scale=0.22]{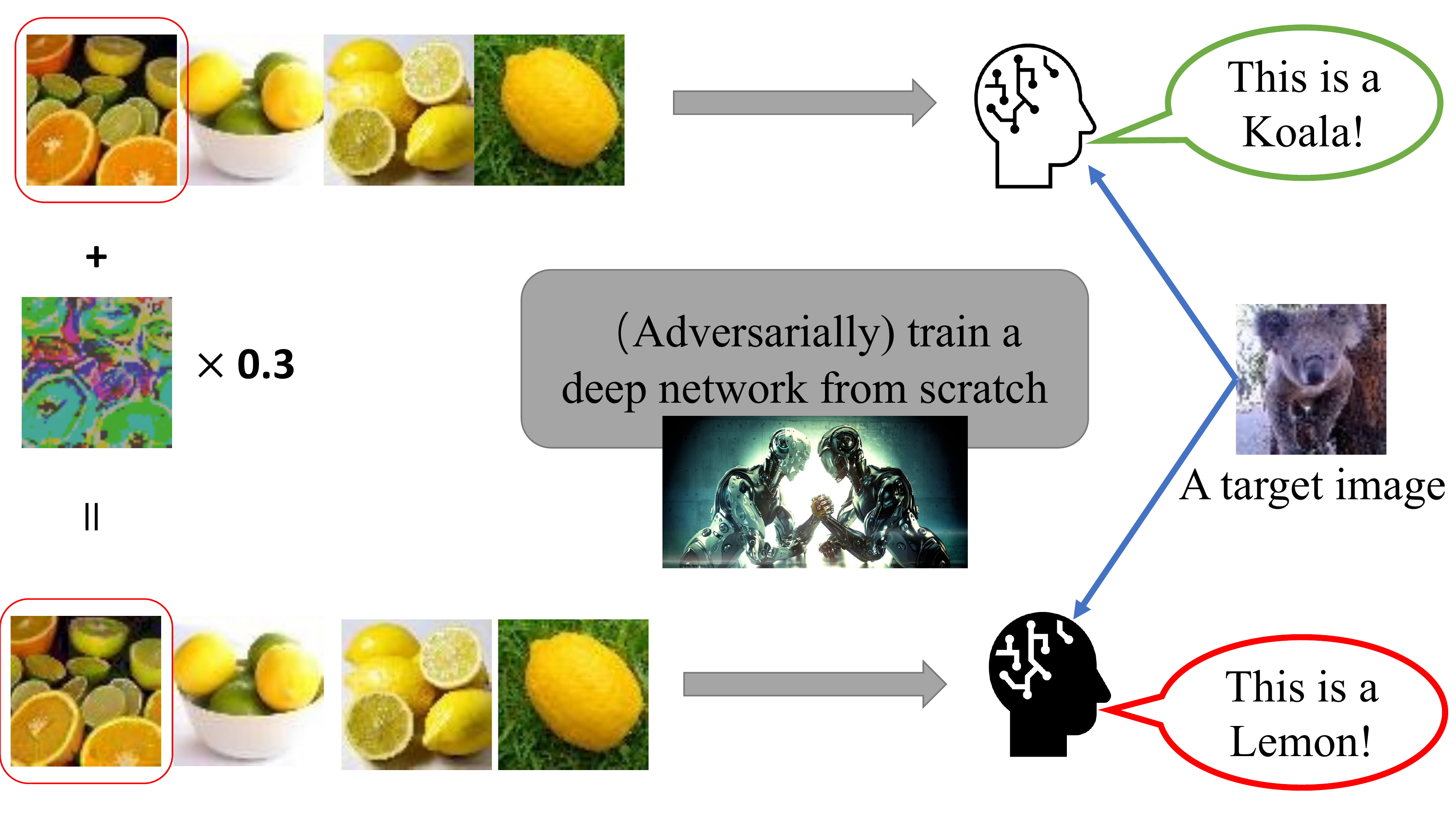}
		\label{fig:illustration_tar}
	}
	\hspace{8mm}
	\subfigure [Clean-label untargeted attack]{
		\includegraphics[scale=0.22]{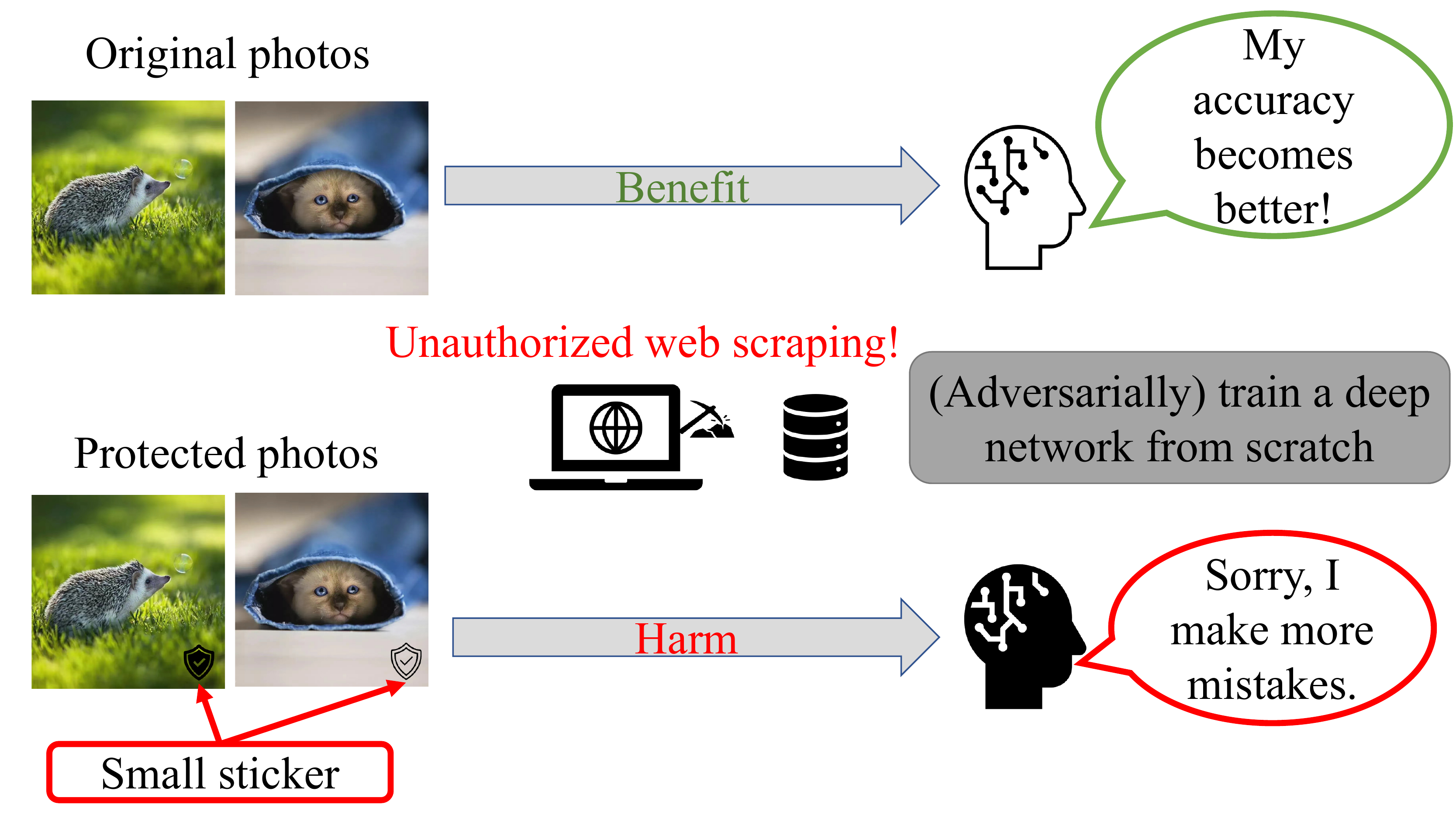}
		\label{fig:illustration_untar}
	}
	\vspace{-2mm}
	\caption{
\textbf{Figure~\ref{fig:illustration_tar}} illustrates the clean-label targeted attack, where the attacker aims to control the behavior of the deep network on a specific target image only. The attacker modifies a tiny fraction of the training data, which appears unmodified and labeled correctly. A learner then trains a network from scratch with this modified dataset. The attacker's minor modification only makes the network output the attacker-specified label on the specific and unperturbed target image without affecting predictions on other input data.\\
In \textbf{Figure~\ref{fig:illustration_untar}}, the clean-label untargeted attack is shown, where the attacker attaches tiny stickers to publicly released images. These stickers signal to the unauthorized data collectors that those images are prohibited from being collected, and otherwise, they will significantly harm the learner, even if the learner employs adversarial training to train a model. This attack aims to minimize both natural generalization and adversarial robustness of the learner with various sizes of perturbation radius $\epsilon_0$.
}
	\vspace{-4mm}
\end{figure*}

In the clean-label targeted attack, the attacker's goal is to gain control over the classifier~\cite{geiping_witches_2021}, even if it has been trained with AT.  
Figure~\ref{fig:illustration_tar} illustrates how the attacker imperceptibly modifies a small portion of the training data, thereby breaching AT's integrity on a specific data point.

In the clean-label untargeted attack, the attacker's objective is to harm the classifier's overall performance. 
As shown in Figure~\ref{fig:illustration_untar}, the attacker attaches a \textit{sticker} to each publicly released data point, which can significantly degrade the classifier's overall performance. These stickers can also signal the prohibition of unauthorized data collection~\cite{huang2021unlearnable,fu2022robust}. 

Admittedly, we cannot claim our proposed attacks are almighty effective since the learner can adapt its behavior in response to the proposed attacks, which is an endless game between an attacker and a learner. However, our proposed attacks answer a scientific question about how the \textit{generalization} and the \textit{robustness} obtained by AT can get affected when encountering the worse-case corrupted training data. 
In other words, we conduct a worst-case analysis of AT's stability with respect to small changes in the training set. By doing so, we shed light on the potential risks of using AT and provide insights for future research.

\section{Background}

\textbf{Notation.} A training set $S = \big\{ (x_1,y_i), ..., (x_n,y_n)\big\}$ is a finite set of data-label pairs in $\mathcal{X} \times \mathcal{Y}$, i.e., a set of labeled data points.
A learning algorithm (the learner) denoted as $\A$ takes $S$ as input and outputs a classifier $f$ , i.e., $\A: S \to f$ and $f: \mathcal{X} \to \mathcal{Y}$.  
The purpose of $\A$ is to output the classifier $f$ with the hope of incurring the minimum loss that is measured by $L$. The measure $L$ could have different criteria, e.g., \textit{natural generalization loss} $  L_{\nat} = \mathbb{E}_{p(x,y)} [\ell(f(x),y )]$ or \textit{robust generalization loss}  $L^{\epsilon}_{\rob} = \mathbb{E}_{p(x,y)}  [\max_{\xadv \in \epsball[x]} \ell(f(\xadv), y)] $, in which $\ell$ is a loss function, i.e., $\ell: \mathcal{X} \times \mathcal{Y} \to \R_{+} $, and $\epsball[x]$ is a closed norm ball of radius $\epsilon >0$ centered at $x$. Note that $L^{\epsilon=0}_{\rob}  = L_{\nat}$.
The robust learner $\A_{\epsilon}$ performs AT, whose purpose is minimizing both $L_{\nat}$ and $L^{\epsilon}_{\rob}$, which corresponds to AT's two purposes---improving both natural generalization and adversarial robustness. 

\subsection{Adversarial Training (AT)} Let $(\cX,d_\mathrm{\infty})$ be the input feature space $\cX$ with the infinity distance metric $d_{\inf}(\bx,\bxprime)=\|\bx-\bxprime\|_\infty$, and specify the $\ell_{\infty}$ closed ball $\epsball[\bx] = \{\bxprime \in \cX \mid d_{\inf}(\bx,\bx')\le\epsilon\}$.
Given a dataset $S$, where ${x} \in \cX$ and $y \in \cY =  \{0, 1, ..., C-1\}$, 
the robust learner $\A_{\epsilon}$ performs $\epsball$-AT aiming at
\begin{equation}
	\label{madry_adversarial_training}
	f_{\epsilon} = \argmin_{f\in\cF} \frac{1}{n}\sum_{i=1}^n \left\{\max_{\bxtidle \in \epsball[\bx_i]} \ell(f(\bxtidle),y_i)\right\},
\end{equation}
where $\bxtidle$ is the adversarial data within the $\epsilon$-ball centered at ${x}$. 
Madry et al. (2018)~\cite{Madry_adversarial_training} approximately solved Eq.\eqref{madry_adversarial_training} via the alternative optimization of an adversarially robust model $f_{\epsilon}$, with one step maximizing the loss to find the adversarial data $\xadv$ and one step minimizing the loss on the generated adversarial data $\xadv$ w.r.t.~model parameters.
There are many AT's variants on specifying different types of norm ball $\cB$, such as spatial AT~\cite{Chaowei_iclr18_deformation} and $\ell_{2}$-norm AT~\cite{Madry_adversarial_training}.
In this paper, we focus on $\ell_{\infty}$-norm AT and leave other AT variants to future explorations.

\subsection{Poisoning Attacks}
In poisoning attacks~\cite{Biggio_poisoning_SVM_ICML2012,Biggio_adv_label_noise}, attackers slightly modify the benign training dataset $S$ to its close counterpart $S'$, which can significantly affect the learner $\A$ w.r.t. its purpose measure $L$.
$L$ is commonly fixed to natural generalization loss $L_{\nat}$. 
Various closeness measures $d(S', S)$ induce different settings of poisoning attacks. 

$S$ and $S'$ could differ by labels $y$. For example, label-flipping set $S'$ significantly increases natural generalization loss $L_{\nat}$ of support vector machines~\cite{Biggio_adv_label_noise,xiao2012adversarial,zhao2017efficient,zhang2017game}, graph neural networks~\cite{zhang2020adversarial} and federated learning~\cite{fung2018mitigating,cao2019understanding,tolpegin2020data}.

$S$ and $S'$ could differ by the input features $x$. 
For example, targeted clean-label poisoning attackers can make human-imperceptible modifications of a part of training images, which dramatically subverts the model's predictions of a test image to an attacker-appointed label~\cite{Shafahi_nips18,huang_matapoison_nips20,geiping_witches_2021}.
Besides, untargeted clean-label poisoning attacks synthesize the crafted human-imperceptible noise into the input data, significantly increasing $L_{\nat}$ of the deep learning models~\cite{shen2019tensorclog,feng_nips19}. Therefore, this type of poisoning attack claims the benefits of data by discouraging personal data from being freely exploited by machines~\cite{zhiqi_privacy,huang2021unlearnable}. 
Furthermore, backdoor attackers~\cite{Gu_acess19,carlini2022poisoning,turner2019label} inject (visibly or invisibly) tiny ``trojans'' into the input data in the training phase, and the ``trojans'' are subsequently invoked at the inference phase. However, modifying the test data is out of this paper's scope. For the discussions between adversarial and backdoor robustness, please refer to~\cite{weng_nips20_backdoor_adversarial,gao2022effectiveness}. 

$S$ and $S'$ could differ by a single point $(x,y)$. Koh and Liang (2017)~\cite{Koh_inference_ICML17} leveraged the influence function to compute the influence of $L_{\nat}$ of ST on removing a particular training data point. Then, Koh et al. (2022)~\cite{koh_MLJ} and Fang et al. (2020)~\cite{fang20_poisoning_rec} leveraged the influence functions to construct poisoning attacks.

Different from prior work considering $L_{\nat}$ only, we study a more challenging attack (training-phase poisoning attack) affecting both $L_{\nat}$ and $L^{\epsilon}_{\rob}$.
We show two examples of clean-label attacks, where the attacker can modify input features without touching labels.  Besides, this work considers a more challenging from-scratch-training AT and a typical case of $\ell_{\infty}$-norm AT. 
We leave poisoning the transfer learning~\cite{Shafahi_nips18,Polytope_EuroSP,zhu_transfer_poison}, other types of AT (such as spatial AT~\cite{Chaowei_iclr18_deformation} and $\ell_{2}$-norm AT~\cite{Madry_adversarial_training}), and label-flipping attacks, and influence functions on AT to future explorations.

\section{Poisoning Attacks Against AT}

In this section, we introduce two novel poisoning attack strategies against AT, namely the \textit{clean-label targeted attack} and the \textit{clean-label untargeted attack}. In subsequent sections, we assume that the robust learner uses $\epsballzero$-AT and that the attacker generates poisoned data with perturbations bounded by a radius of $\epsilon$.\footnote{Throughout the paper, $\epsilon_{0}$ is the parameter associated with the learner, while $\epsilon$ is the parameter associated with the attacker.}

\subsection{Threat Model}
\textbf{Attacker's goal.}
In the clean-label targeted attack, the attacker's objective is to control the behavior of the classifier $f_{\epsilon_{0}}$ (returned by a robust learner $\A_{\epsilon_{0}}$) on a specific test data point $(\xtar,y_{\tar})$ without degrading overall classification performance, which makes this attack insidiously hard to detect~\cite{Shafahi_nips18,huang_matapoison_nips20,geiping_witches_2021}. The attacker wins if the learned classifier $f_{\epsilon_{0}}(\xtar) = y_{\adv}$ or $f_{\epsilon_{0}}(\xadv_{\tar}) = y_{\adv}$, where $y_{\adv} \neq y_{\tar}$ is the attacker-specified label, and $\xadv_{\tar} \in \epsballzero[\xtar]$.

In the clean-label untargeted attack, the attacker aims to significantly degrade natural generalization and adversarial robustness of robust learner $\A_{\epsilon_{0}}$ with various size of $\epsilon_{0}$, i.e., maximizing  both $L_{\nat}$ and $L^{\epsilon_{0}}_{\rob}$ on all test data. 
This attack protects the data from unauthorized collection that enhances the model's utility (targeting at minimizing $L_{\nat}$)~\cite{zhiqi_privacy,shen2019tensorclog,shan2020protecting,huang2021unlearnable,radiya-dixit2022data,fu2022robust} or robustness (targeting at minimizing $L^{\epsilon_{0}}_{\rob}$).

\noindent\textbf{Attacker's knowledge.}
The attacker can get access to the entire training data and know the learner $\A_{\epsilon_{0}}$ will perform $\epsballzero$-AT with the $\ell_{\infty}$-norm to enhance some robustness of the model. However, the attacker is unaware of the network structure and weight initialization that the learner $\A_{\epsilon_{0}}$ uses in AT. 
\\
\noindent\textbf{Attacker's capability.}
The attackers cannot control the labeling of training data but can modify their input features without changing the semantic meanings.
In the clean-label targeted attack, the attacker can imperceptibly modify a small fraction of the training data before training.
In the clean-label untargeted attack, the attacker can add a visible but tiny \textit{sticker} on each training data point signal to the data collector that this data point is prohibited from unauthorized collection.


\subsection{Clean-label Targeted Attack Strategy}
We consider a challenging poisoning attack against the from-scratch-training AT on deep neural networks (DNNs). 
Compared with poisoning the linear classifiers~\cite{Biggio_poisoning_SVM_ICML2012,xiao2015feature} and the fine-tuning process~\cite{Shafahi_nips18}, it has been proven more challenging to poison the from-scratch-training DNNs~\cite{munoz2017towards,Shafahi_nips18,huang_matapoison_nips20,geiping_witches_2021,schwarzschild2021just}. Besides, it has been shown AT inherently resists the corrupted training data to some extent~\cite{sanyal2020benign,tao_delusive}, which suggests that it is even more challenging to poison AT.

We adopt the idea of \textit{gradient matching}~\cite{zhao2021dataset,geiping_witches_2021}. Geiping et al. (2021)~\cite{geiping_witches_2021} fixed a single pretrained model $f$ (parameterized using $\theta$) by ST and matched the model gradient of the poisoned data with those of targeted data with attacker-specified label $y_{\adv} \neq y_{\tar}$, i.e., 
\begin{equation}
	\label{x_poi_gen}
	\argmin_{x_{\poi} \in \epsball[x]} \ML \Big(\nabla_{\theta} \ell\big( f(x_{\poi}), y \big) ,\nabla_{\theta} \ell \big(f(\xtar), y_{\adv}) \big)   \Big), 
\end{equation} 
where $x_{\poi}$ is the generated poisoned variant of its natural counterpart $x$ in the training set, $\xtar$ is a selected data point in the test set, and $\ML(\cdot, \cdot)$ is a matching loss, e.g., the cosine similarity loss  $\ML(\Vec{a}, \Vec{b}) = \frac{\Vec{a} \cdot \Vec{b}}{||\Vec{a}|| ||\Vec{b}||}$.
The target and the poisoned gradients are aligned in the same direction so that the poisoned data can mimic the gradient of the targeted data during the training. Consequently, it achieves the state-of-the-art targeted poisoning attacks against the from-scratch-training DNNs.

However, AT does not directly learn from natural data $x$ but from its adversarial variants $\xadv \in \epsball[x]$, which inevitably make $x_{\poi}\in \epsball[x]$ (generated by Eq.\eqref{x_poi_gen}) ineffective (see Section~\ref{sec:experiment} for validation). 
Therefore, when the attacker is unaware of the learner using AT, the crafted poisoned data~\cite{munoz2017towards,Shafahi_nips18,huang_matapoison_nips20,geiping_witches_2021,schwarzschild2021just} are effective in ST but not in AT at all.

To make the poisoned data effective in $\mathcal{B}_{\epsilon_{0}}$-AT (performed by a robust learner $\A_{\epsilon_{0}}$), we propose that the pretrained model should choose a robust one $f_{\epsilon_{0}}$ that is returned by $\A_{\epsilon_{0}}$ on clean set $S$, and the gradients of the adversarial variants of the poisoned data should match those of targeted data. Therefore, we have the following objective for generating the poisoned data as follows.
\begin{align}
	\label{x_poi_adv_gen}
	\argmin_{x_{\poi} \in \epsball[x]} &\ML \Big(\nabla_{\theta} \ell\big(  f_{\epsilon_{0}} (\tilde{x}_{\poi}), y \big) ,\nabla_{\theta} \ell \big(f_{\epsilon_{0}}(\xtar), y_{\adv}) \big)   \Big),  \\ 
	\label{x_poi_adv}
	\tilde{x}_{\poi} &= \argmax_{\tilde{x}_{\poi} \in \mathcal{B}_{\epsilon_{0}}[x_{\poi}] } \ell\big(  f_{\epsilon_{0}} (\tilde{x}_{\poi}), y \big).
\end{align}
The generated poisoned data will be more effective with larger $\epsilon$. Notably, only $\epsilon > \epsilon_{0}$ can potentially generate the poisoned data that mislead the robust learner $\A_{\epsilon_{0}}$ on $\xtar$, but any size of $\epsilon > 0$ has a potential of poisoning $\A_{\epsilon_{0}}$ on adversarial variants of $\tilde{x}_{\tar} \in \mathcal{B}_{\epsilon_{0}}[\xtar]$.

\begin{algorithm}[tp!]
	\caption{Clean-label targeted poisoning attack against adversarial training}
	\label{alg:clean-label}
	\begin{algorithmic}
		\STATE {\bfseries Input:}  Clean training set $S$ of size $n$. Prior knowledge of $\A_{\epsilon_{0}}$ learner performing $\mathcal{B}_{\epsilon_{0}}$-AT. Specify a small portion $\rho$ (from the same base class) of the training data $\{(x_1, y), ...,(x_m, y)\} \subset S$ and $m \ll n$. Specify a target and its label $(\xtar, y_{\adv})$, $y_{\adv}\neq y_{\tar}$, and perturbation radius $\epsilon$.
		\STATE {\bfseries Output:}  Poisoned training set $S'$ containing $m$ invisibly poisoned data and each $x_{\poi} \in \epsball[x]$.
		\STATE {\bfseries Step 1:} Mimic the robust learner $\A_{\epsilon_{0}}$ on $S$ to obtained a learned $f_{\epsilon_{0}}$. 
		\STATE{\bfseries Step 2:} Compute and optimize the averaged loss in Eq.\eqref{learning_obj} over $m$  data  and return $x_{\poi}$ of each $x$.
	\end{algorithmic}
\end{algorithm} 

Directly solving Eq.\eqref{x_poi_adv_gen} incurs inefficient bi-level optimization; therefore, we seek approximations to enhance its computational efficiency. Based on the triangle inequality (i.e., $\ML(\Vec{a}, \Vec{b}) \leq \ML(\Vec{a}, \Vec{c}) + \ML(\Vec{b}, \Vec{c})$), we convert Eq.\eqref{x_poi_adv_gen} into 
\begin{equation}
	\begin{aligned}
		\label{x_poi_two_terms}
		\fontsize{8}{8}
		\argmin_{x_{\poi} \in \epsball[x]} &\Big\{ \ML \Big(\nabla_{\theta} \ell\big(  f_{\epsilon_{0}} (\tilde{x}_{\poi}), y \big) ,\nabla_{\theta} \ell \big(f_{\epsilon_{0}}(x_{\poi}), y) \big)   \Big)\\
		+ \ML &\Big(\nabla_{\theta} \ell \big(f_{\epsilon_{0}}(\xtar), y_{\adv}\big) , \nabla_{\theta} \ell \big(f_{\epsilon_{0}}(x_{\poi}), y) \big) \Big)  \Big\}, 
	\end{aligned}
\end{equation} where the first term that is involved in the difficult bi-level optimization matches the adversarial variant $\tilde{x}_{\poi}$ with its natural counterpart $x_{\poi}$. To minimize the first term, we can enforce $\tilde{x}_{\poi} \approx {x}_{\poi}$.
Note that by Eq.\eqref{x_poi_adv}, $\tilde{x}_{\poi}$ maximizes $\ell$ within $\mathcal{B}_{\epsilon_{0}}[x_{\poi}]$ distance of $x_{\poi}$. $\tilde{x}_{\poi}$ will not be different from $x_{\poi}$ if $x_{\poi}$ makes $\ell$ largest already. Based on this heuristic, we approximate Eq.\eqref{x_poi_two_terms} by optimizing 
\begin{equation}
	\begin{aligned}
		\label{learning_obj}
		\argmin_{x_{\poi} \in \epsball[x]} &\Big\{ - \lambda \ell\Big(f_{\epsilon_{0}} ({x}_{\poi}), y \Big) \\
		+ \ML &\Big(\nabla_{\theta} \ell \big(f_{\epsilon_{0}}(\xtar), y_{\adv}\big) , \nabla_{\theta} \ell \big(f_{\epsilon_{0}}(x_{\poi}), y) \big) \Big) \Big\},
	\end{aligned}
\end{equation}
where $\lambda$ is a hyperparameter that balances the optimization of the two terms. Optimizing Eq.\eqref{learning_obj} is computationally more efficient than Eq.\eqref{x_poi_adv_gen}, which induces our Algorithm~\ref{alg:clean-label} of the clean-label targeted poisoning attack against AT. 

The key take-away message is knowing the enemy. Being aware of the learner that performs the AT, the attacker can choose a robust pre-trained model to guide the generation of the poisoned data.
In the following subsection, we show another facet of knowing the enemy: being aware of the learner performing $\epsballzero$-AT with various size of $\epsilon_{0}$, the attacker can specify a different type of $\epsball'$ that bounds the generation of the poisoned data. For example, being aware of $\mathcal{B}$ being an $\ell_{\infty}$-norm ball, the attacker could specify a different $\mathcal{B}'$ being an $\ell_{0}$-norm ball.

\subsection{Clean-label Untargeted Attack Strategy}
To prevent unauthorized data collection, the clean-label untargeted attacker modifies the training data $S$ to $S'$ that harms the standard learner $\A$~\cite{huang2021unlearnable,shen2019tensorclog,shan2020protecting,zhiqi_privacy,fowl_adv_strong_poi,yuan_Nerual_Tangent_attack}, which is no longer effective in the robust learner $\A_{\epsilon_{0}}$~\cite{fu2022robust}. 
Fu et al. (2022)~\cite{fu2022robust} conducted a pioneer study on crafting the \textit{robust error-minimizing} (REM) noise (invisible to humans) to harm the robust learner $\A_{\epsilon_{0}}$.  The essential idea is \textit{knowing the enemy}: being aware of the robust learner performing $\mathcal{B}_{\epsilon_{0}}$-AT, the attacker trains a robust error minimizing noise generator $g_{\epsball} $ as follows. 
\begin{equation}
	\label{eq:train_gen}
	g_{\epsball} = \argmin_{g \in\cF} \frac{1}{n}\sum_{i=1}^n \left\{    \min_{\bx' \in \epsball[x_i]}   \max_{\bxtidle \in {\mathcal{B}_{\epsilon_{0}}} [\bx']} \ell(g(\bxtidle),y_i)\right\},
\end{equation} where 
$\epsilon_{0}$ is the perturbation radius of the targeted robust learner $\A_{\epsilon_{0}}$, $\epsilon$ is the perturbation radius that bounds REM noise, and  
$\epsilon$ should be set larger than $\epsilon_{0}$; $\bxtidle$ is the intermediate data that composites the natural data $x_i$, $\epsilon_{0}$-noise and $\epsilon$-noise. 

Then, the poisoned data are generated by $x_{\poi} = \argmin_{x_{\poi} \in \epsball [\bx]} \{  \ell(g_{\epsball}(x_{\poi}), y )   \}$. Please refer to \cite{fu2022robust} for the detailed implementations of Eq.\eqref{eq:train_gen}.

However, the min-min-max optimization in Eq.\eqref{eq:train_gen} incurs the heavy computation costs and the training instability. 
Furthermore, it is worth noting that once the poisoned data are released, the attacker could not modify the data any further. Therefore, when the robust learner increases $\epsilon_{0}$ to $\epsilon$, the aforementioned method is less and less effective (see Section~\ref{sec:experiment} for the validation).

To ease the above issues, we provide an alternative facet of \textit{knowing the enemy}: the attacker can specify a different type of $\epsball'$ (different from the learner's $\epsball$) that bounds the generation of the poisoned data. 
Specifically, we add the human-visible stickers to the released data, which can signals to those unauthorized data collectors that the data are prohibited from being collected and, otherwise, will significantly harm the learner. Therefore, those stickers could have a sense of signaling ownership. 

The stickers should be visually small and not affect the normal usage. Therefore, the stickers can be bounded by the $\ell_{0}$-norm (denoted as $\epsball'$) that measures the Hamming distance between the two images~\cite{shamir2019simple}.
Specifically, we only allow the attacker to change a small patch of the image, but each pixel in the small patch can get changed arbitrarily.
To this end, we learn a generator $g_{\epsball'}$ to generate the stickers:  
\begin{equation}
	\label{eq:train_gen_privacy}
	g_{\epsball'} = \argmin_{g \in\cF} \frac{1}{n}\sum_{i=1}^n \left\{ \min_{\bx' \in \epsball'[x_i]}   \ell(g(\bx'),y_i)\right\},
\end{equation} 
where $x'$ is the image attached with the sticker. We alternatively optimize the sticker and the parameters of generator $g_{\epsball'}$. 
Algorithm~\ref{alg:clean-label-untar} shows our clean-label untargeted poisoning attacks in detail. 

\begin{algorithm}[tp!]
	\caption{Clean-label untargeted poisoning attack against adversarial training}
	\label{alg:clean-label-untar}
	\begin{algorithmic}
		\STATE {\bfseries Input:}  Clean training set $S$. Prior knowledge of the robust learner $\A_{\epsilon_{0}}$ specifying $\mathcal{B}$ as $\ell_{\infty}$-norm with various size of ${\epsilon_{0}}$. Randomly initialize the sticker (denoted as ``Patch'')  valued between $[0,1]$. The ``MASK'' valued at $\{0,1\}$ specifies the shape and the position of the sticker. 
		\STATE {\bfseries Output:}  Poisoned training set $S'$ that prevents the unauthorized collection of the data.
		\STATE {\bfseries Step 1:} Learn a sticker generator $g_{\epsball'}$ 
		\FOR{Epoch  $e= 1$, $\dots$, $E$}
		\FOR{each data $x$ (or batch) in the training set $S$}
		\STATE{$x' = x \odot (1-\text{MASK}) + \text{Patch} \odot \text{MASK}$, where $\odot$ is Hadamard product.}
		\STATE{Fix the generator $g_{\epsball'}$ and optimize and update ``$\text{Patch}$'' via minimizing  $\ell(g_{\epsball'}(\bx'),y_i)$.}
		\STATE{Fix the ``Patch'' and optimize and update $g_{\epsball'}$ via minimizing  $\ell(g_{\epsball'}(\bx'),y_i)$.}
		\ENDFOR
		\ENDFOR
		\STATE{\bfseries Step 2:} Use the generator to attach sticker to each data $x$.
		\STATE{For each data $x \in S$, output $\bx' \in \epsball'[x]$  via minimizing $\text{``Patch''}$ on $\ell(g_{\epsball'}(\bx'),y_i)$}.
	\end{algorithmic}
\end{algorithm} 
Algorithm~\ref{alg:clean-label-untar}  does not employ the robust model for generating poisoned data; therefore, we cannot expect the sticker consistently outperforms REM, especially under the small $\epsilon_{0}$ regime. 
However, when $\epsilon_{0}$ gets larger, REM inevitably becomes less and less effective, but we can expect sticker is still effective. 
It is worth noting that as long as the stickers can degrade the learner's performance even to a small extent, we can firmly expect the discouragement effect that the data being collected in an unauthorized way.

\subsection{Discussion of the Arms Race}
\label{sec:arm-race}
Security is a reactive arms race~\cite{biggio2018wild}, where both the attacker and the learner can adapt their behavior in response to each other.
The side with more knowledge typically has the advantage.
In this paper, we show that the attacker can anticipate that the learner will perform $\ell_{\infty}$-AT and can design adaptive attack strategies accordingly.
However, we should avoid claiming that any attack strategy is almighty effective because the learner can also anticipate the attacker and design a corresponding defense. This is an endless game of one-upmanship. 

We believe that the attacker can significantly increase the cost for the learner. To do this, the attacker could take a proactive approach by a) identifying the potential learning strategies that the learner may employ, 
b) designing adaptive attacks for each learning strategy, and c) revising and repeating this process if necessary. 
For example, the attacker could output poisoned data set $S'$, where each small portion is poisoned using different strategies such as our method, REM and flipping-label attacks. 
To counteract these various poisons, the learner would have to incorporate multiple defensive learning strategies simultaneously, which would require significant computational resources.

\begin{figure*}[tp!]
	\vspace{0mm}
	\subfigure[CIFAR-10]{
		\begin{minipage}[t]{0.4\textwidth}
			\centering
			\includegraphics[scale=0.7]{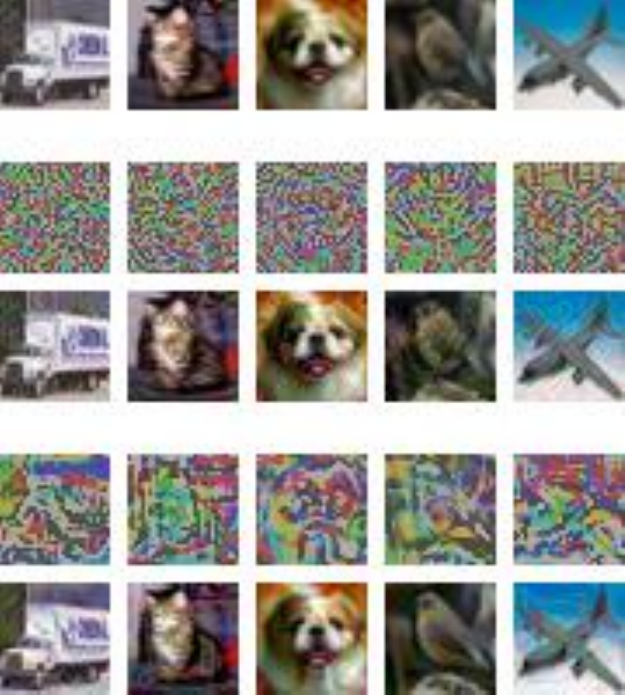}
		\end{minipage}
	}	
	\centering
	\subfigure[Tiny ImageNet]{
		\begin{minipage}[t]{0.5\textwidth}
			\centering
			\includegraphics[scale=0.4]{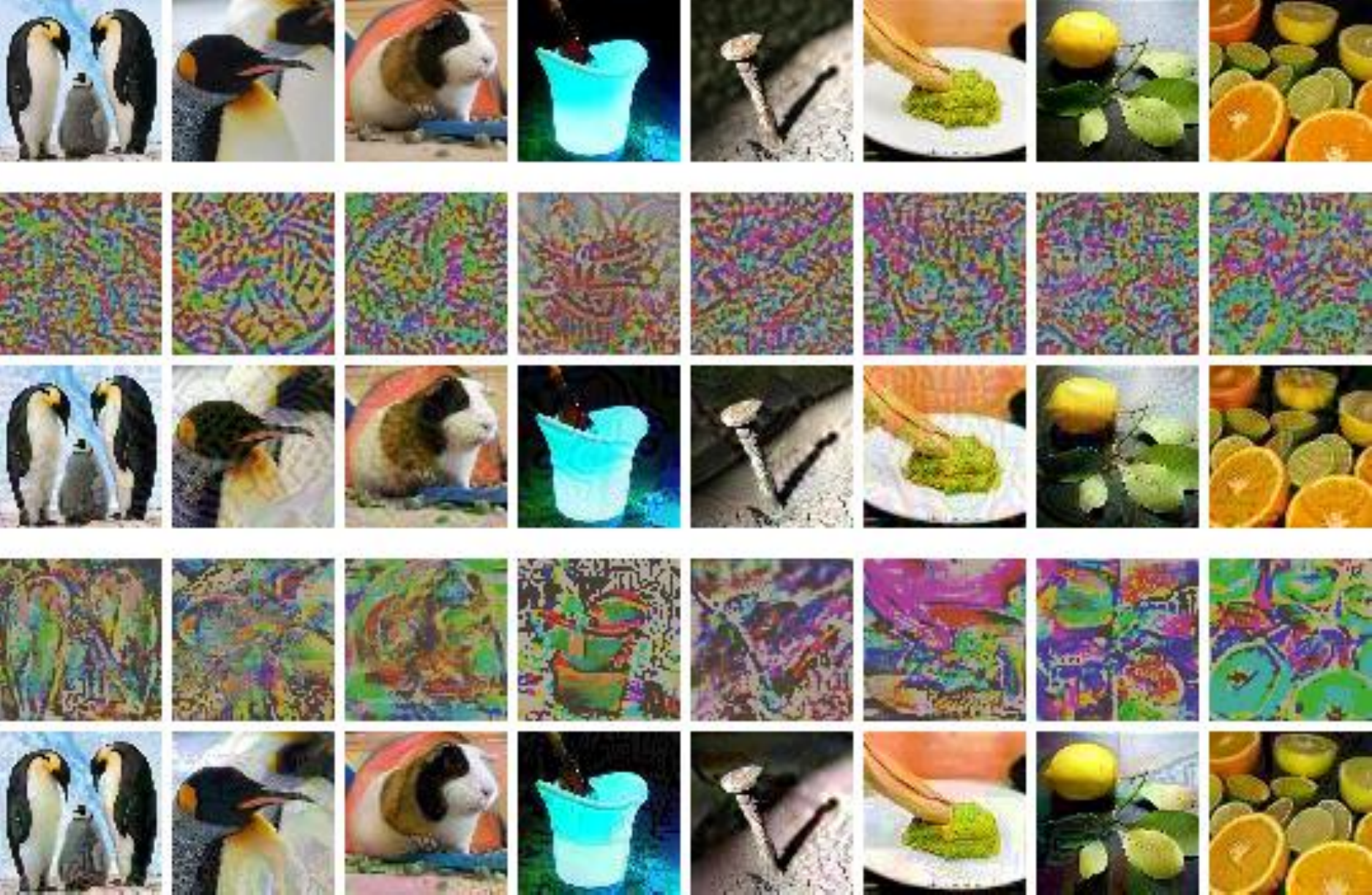}
		\end{minipage}
	}
	\vspace{-2mm}
	\caption{provides a visualization of clean-label targeted poisoning attacks carried out by Witches' Brew (WB) and our proposed method (by Algorithm~\ref{alg:clean-label}), respectively. The top row shows unperturbed natural images, while the second and third rows depict the imperceptible noise (scaled by 3 times for visualization) and the poisoned images by WB. Similarly, the fourth and fifth rows depict the imperceptible noise (scaled by 3 times for visualization) and the poisoned images generated by our method. All poisoning noises are bounded by $\ell_{\infty}$-norm with $\epsilon = 16/255$.}
	\label{fig:samples}
        \vspace{-3mm}
\end{figure*}

\section{Experiment}
\label{sec:experiment}

\textbf{Hardware Setup.} All experiments are conducted using NVIDIA GEFORCE RTX 3090 GPUs and Intel Xeon Gold 6248R CPUs. Specifically, we use one GPU for each experiment on CIFAR-10~\cite{krizhevsky2009learning_cifar10}, CIFAR-100~\cite{krizhevsky2009learning_cifar10}, and Tiny ImageNet~\cite{li_feifei_tinyimagenet}. For the experiments on ImageNet Subset~\cite{deng2009imagenet}, we use three GPUs.

\subsection{Clean-label Targeted Attacks Against AT}
\label{sec_exp:clean-label-tar}
\noindent\textbf{Data augmentation.}
We conduct experiments on CIFAR-10 and Tiny ImageNet datasets. 
We normalize all images to $[0,1]$ and apply random translation, random crop, and random horizontal flip when the robust learner uses AT. However, we do not apply any data augmentation when generating the poisoned data in Algorithm~\ref{alg:clean-label}. 

\noindent\textbf{Robust learner.}
We employ standard AT, i.e., $\ell_{\infty}$-norm AT and use the $\ell_{\infty}$ projected gradient descent (PGD) method~\cite{Madry_adversarial_training} to generate adversarial examples for both training and testing. All PGD methods have random initialization enabled. We set the step number and step size of PGD as $10$ and $\epsilon_{0}/4$, respectively. Furthermore, we train all the models with a batch size of $128$ and a weight decay factor of $0.0005$. 

\noindent\textbf{Witches' brew (WB) Implementation. }
WB~\cite{geiping_witches_2021} aims to poison ST and then trains a surrogate ResNet-18 model using standard training procedures. The projected ADAM optimizer is used with the step size of 0.1 to update the poisoned samples via Eq.\eqref{x_poi_gen}. \textit{Differentiable data augmentation} techniques, as described in the original paper, are enabled.

\noindent\textbf{Our attacker.}
In our experiments, our attacker aims to poison the AT approach by mimicking the robust learner, obtaining a robust model on the clean set $S$, and generating poisoned data based on this model.

\noindent\textbf{Generation of imperceptibly poisoned data.}
The attacker's target is to control the behavior of the robust learner $\A_{\epsilon_{0}}$ on a targeted data point, where the $\epsilon_{0}$ is fixed at $2/255$. 
We first mimic $\A_{\epsilon_{0}}$ using ResNet-18~\cite{he2016deep} on the clean set $S$ and obtain a robust ResNet-18 (i.e., $f_{\epsilon_{0}}$). On both CIFAR-10 and
Tiny ImageNet, we adversarially train the ResNet-18 for 40 epochs, using SGD with 0.9 momentum and setting the initial learning rate of $0.1$ that is decayed by $10$ three times.

Next, given a clean dataset $S$, we randomly choose $\rho$ portion (from the same base class) of training data, randomly specify a target data point $x_{\tar}$ in the test set, and randomly specify its adversarial label $y_{\adv} \ne y_{\tar}$. We set $\rho = 0.04$ for CIFAR-10 and $\rho=0.005$ for Tiny ImageNet, respectively. 
We choose $\epsilon \in [4/255, 16/255]$ to generate the poisoned data. 
We initialize the poisoned noise using the random Gaussian noise and then use the PGD method with a step size of $0.01$ as an optimizer to update poisoned samples using Eq.\eqref{learning_obj}.
We set the hyperparameter $\lambda$ to $0.01$ for CIFAR-10 and $0.001$ for Tiny ImageNet. 
Additionally, we updated the poisoned samples with a batch size of 512 for CIFAR-10 and 128 for Tiny ImageNet.
For each generation, we use the PGD method to optimize Eq.\eqref{learning_obj} with 250 iterations. 
Then, the poisoned dataset $S'$ contains a small number of the poisoned data.
Later, we will validate later whether the $S'$ can control the $\A_{\epsilon_{0}}$ behavior on $x_{\tar}$. 

\noindent\textbf{Evaluation.}
\begin{table*}[tp!] \footnotesize \centering
\caption{Poison settings for various random seeds on CIFAR-10 (left table) and Tiny ImageNet (right table). \\ Target ID refers to a specific data point identifier in the validation set.}
\label{table:targeted_setting}
    \begin{tabular}{cccc} 
				\toprule
				Base Class & Adversarial Class & Target ID & Random Seed \\
				\midrule
				dog & frog & 8745 & 2000000000  \\
				frog & truck & 1565 & 2100000000 \\
				frog & bird & 2138 & 2110000000  \\
				airplane & dog & 5036 & 2111000000  \\
				airplane & ship & 1183 & 2111100000  \\
				cat & airplane & 7352 & 2111110000  \\
				automobile & frog & 3544 & 2111111000  \\
				truck & cat & 3676 & 2111111100  \\
				automobile & ship & 9882 & 2111111110  \\
				automobile & cat & 3028 & 2111111111  \\
				\bottomrule 
		\end{tabular}
  \hspace{8mm}
    \begin{tabular}{cccc}
				\toprule
				Base Class & Adversarial Class & Target ID & Random Seed \\
				\midrule
				frying pan & nail & 4989 & 1000000000  \\
				CD player & lemon & 9731 & 1100000000 \\
				mashed potato & king penguin & 2533 & 1110000000  \\
				cash machine & parking meter & 2088 & 1111000000  \\
				crane & nail & 5439 & 1111100000  \\
				lakeside & centipede & 5723 & 1111110000  \\
				lemon & guinea pig & 5743 & 1111111000  \\
				baboon & spiny lobster & 1171 & 1111111100  \\
				pole & guacamole & 2465 & 1111111110  \\
				potter's wheel & bucket & 7658 & 1111111111  \\
				\bottomrule 
		\end{tabular}
  \vspace{-2mm}
\end{table*}
\begin{figure*}[tp!]
	\centering
	\includegraphics[scale=0.28]{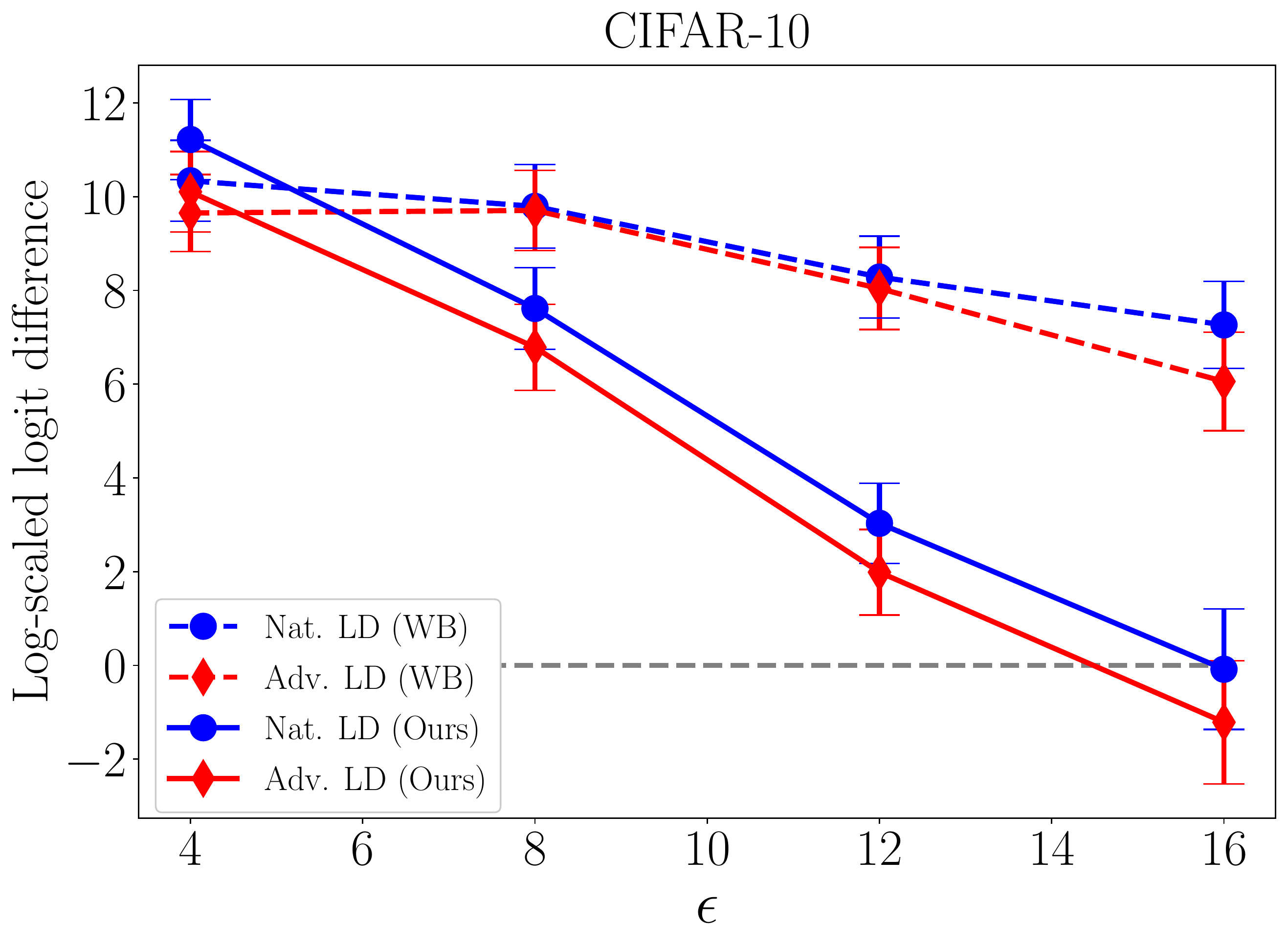}
	\hspace{5mm}
	\includegraphics[scale=0.28]{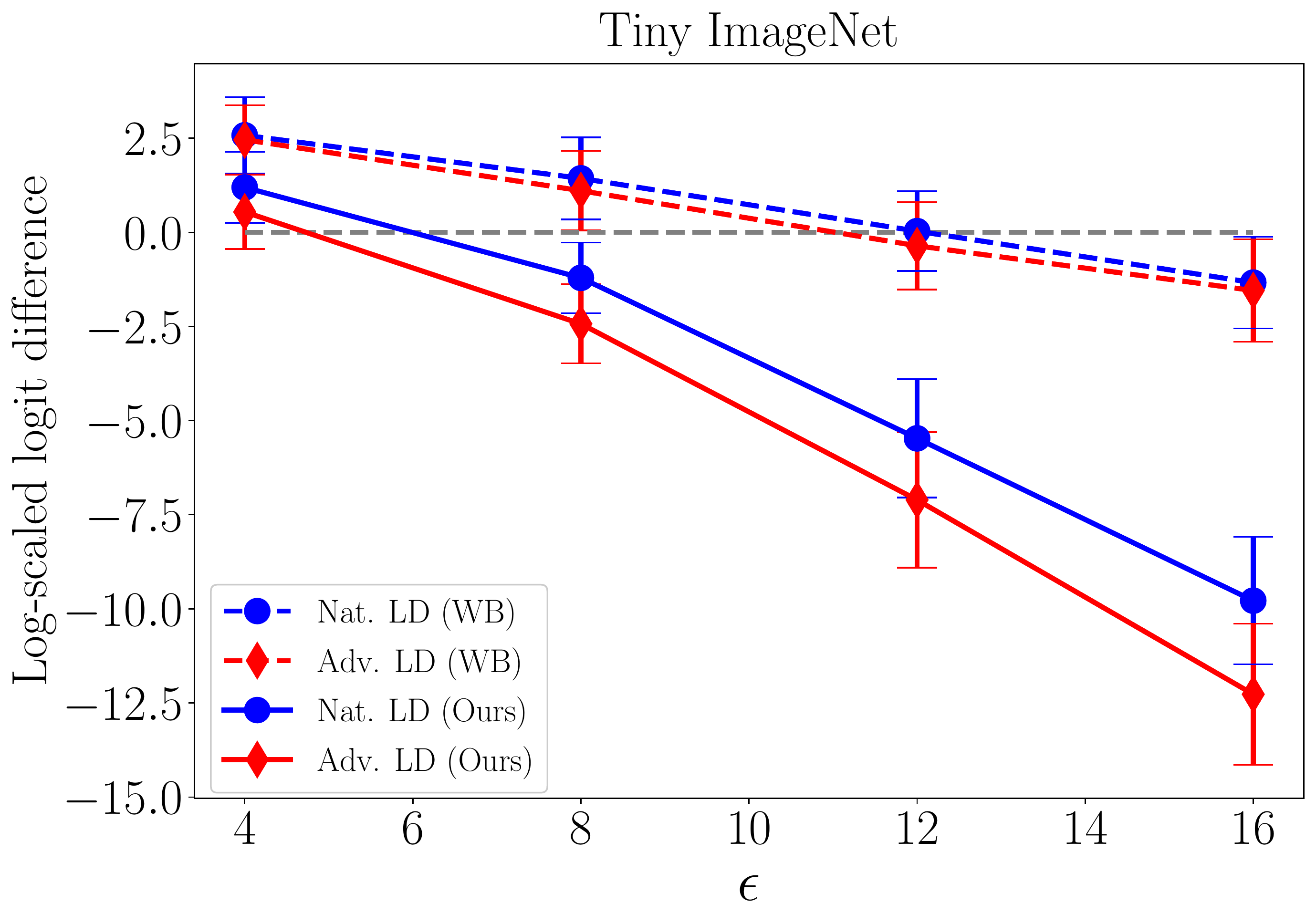} \\
	\vspace{-2mm}
	\caption{Clean-label targeted poison attacks against a robust learner $\A_{\epsilon_{0}=2/255}$ with a different perturbation radius $\epsilon$ by a attacker. The points below the gray dashed line signify the success of the poisoning attack in manipulating the robust learner's behavior on a chosen target.}
	\vspace{-2mm}
	\label{fig:logit_difference}
\end{figure*}
We repeat the above generations of poisoned data ten times, each time with different target data points $x_{\tar}$, adversarial labels $y_{\adv}$ and base classes from which the poisoned training data is drawn (see Table~\ref{table:targeted_setting} for the random seeds used).
For each value of $\epsilon$ in the range $[4/255, 16/255]$, we generate ten poisoned set $S'$. 

To evaluate the effectiveness of the poisoned data, we test $\epsballzero$-AT ($\epsilon_{0} = 2/255$) five times on each poisoned set, using four ResNet-18 models with different initialization and one VGG-11 model. 
We perform AT for 40 epochs, which strikes a balance between computational efficiency and avoiding robust overfitting, as robust overfitting can make poisoned data more effective (see Section~\ref{sec:tar_ablation} for details).

Figure~\ref{fig:logit_difference} compares our method (solid lines) with the state-of-the-art method Witches' Brew (WB~\cite{geiping_witches_2021}, dashed lines) on both CIFAR-10 and Tiny ImageNet, for various values of $\epsilon$ and poison ratios $\rho$. 
We use the log-scaled predictive logit difference (LD), defined as $\big( \log( f_{\epsilon_{0}}^{y_{\adv}} ({\cdot}) )- \log ( f_{\epsilon_{0}}^{y_{\tar}} ({\cdot})) \big)$, to measure the effectiveness of the poisoned data. 
We calculate the Nat. LD on the natural target point $x_{\tar}$, and the Adv. LD on the adversarial target point $\xadv_{\tar} = \argmax_{\xadv_{\tar} \in \epsballzero[x_{\tar}]} \ell(f_{\epsilon_{0}}(\xadv_{\tar}), y_{\tar} )$.
For each value of $\epsilon$, we obtain $10\times5$ LD values for both WB and our method, and report the median LD with standard deviation (error bar).
%
%
%
%
\subsubsection{Main Results of Targeted Poisoning Attacks}
Our experiments show that a robust pre-trained model, as used by our method, is more effective in guiding the generation of poisoned data to fool a robust learner than a standard pre-trained model, as used by Geiping et al. (2021)~\cite{geiping_witches_2021}. Figure\ref{fig:logit_difference} illustrates that the solid lines representing our method are generally below the dashed lines representing Geiping et al.'s method in both CIFAR-10 and Tiny ImageNet experiments.

We also find that poisoning the AT process is harder than poisoning the ST process. In the CIFAR-10 experiment in Figure~\ref{fig:logit_difference}, a larger value of $\epsilon=16/255$ is required to poison the AT process (i.e., logit difference below the gray horizontal line), but Geiping et al. showed that $\epsilon=16/255$ is sufficient to poison ST, and an even smaller value of $\epsilon=8/255$ can poison ST successfully. 


Furthermore, our results suggest that a robust learner with more classes is more vulnerable to poisoning attacks. Our method can successfully poison Tiny ImageNet (200 classes) with $\epsilon=8/255$, while $\epsilon=8/255$ is not always successful in poisoning CIFAR-10 (10 classes). We also used a smaller poison portion $\rho$ in Tiny ImageNet than in CIFAR-10.


Finally, Figure~\ref{fig:samples} provides a comparison of the poisoning noise generated by our method and Geiping et al.'s method. We find that the noise generated by our method contains visually closer semantics and exhibits stronger toxicity against a robust learner. 
%
%
\subsubsection{Ablation Studies}
\label{sec:tar_ablation}
In this section, we conduct ablation studies to fully understand clean-label targeted attacks against AT.

\begin{figure}[tp!]
        \vspace{-0mm}
	\centering
	\includegraphics[scale=0.24]{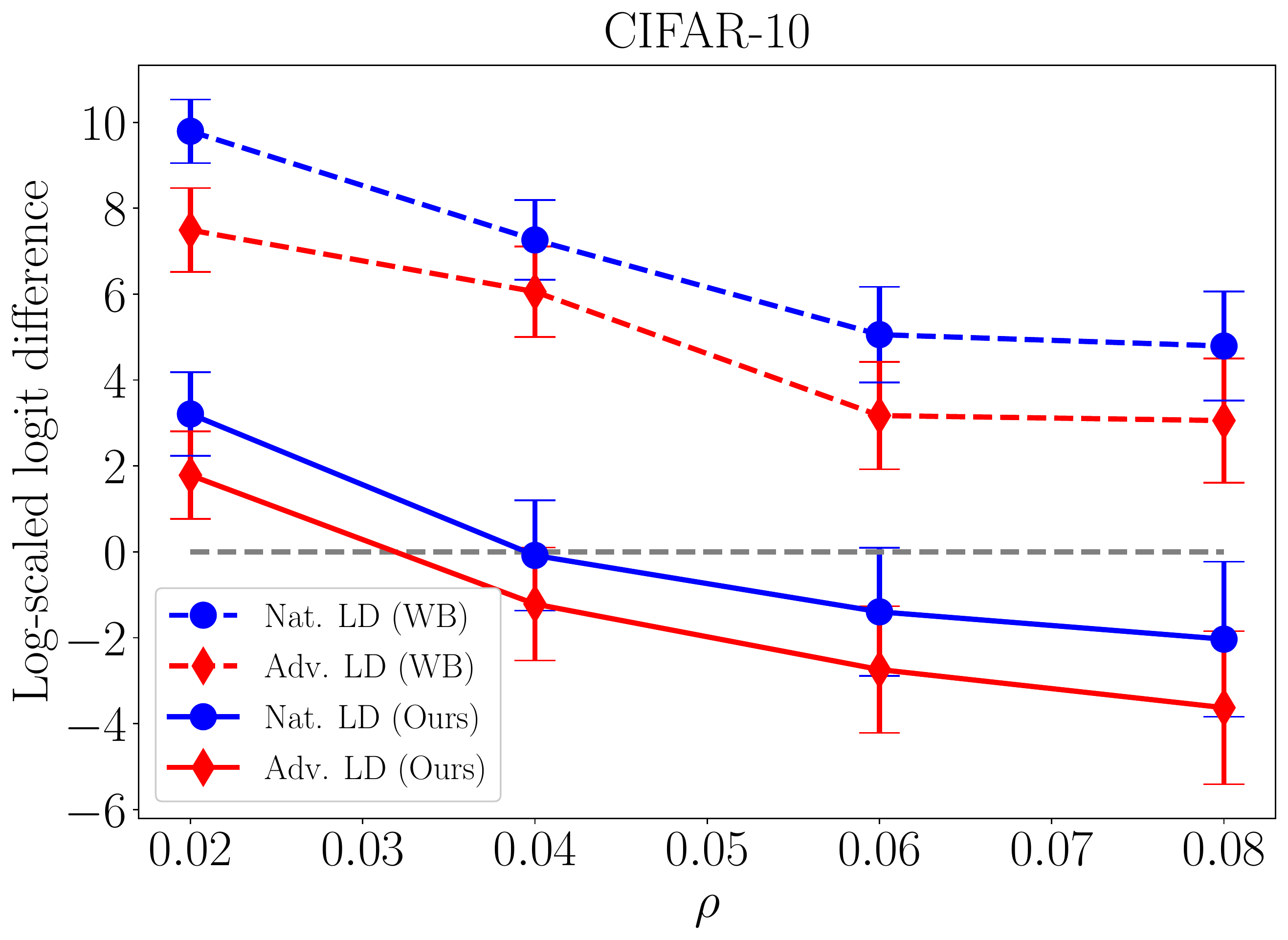} \\
	\vspace{-2mm}
	\caption{Clean-label targeted attacks under various poison budgets $\rho$.}
	\label{fig:budget_logit_difference}
        \vspace{-2mm}
\end{figure}

\noindent\textbf{Effect of different poison budget $\rho$.}
In the left panel of Figure~\ref{fig:logit_difference}, we employ a poison budget $\rho$ of 0.04 in CIFAR-10.
To understand the effect of different poison budgets, we generate poisons with a $\ell_{\infty}$ bound of $\epsilon = 16/255$ and various $\rho$. We keep the same generation and evaluation settings as in Figure~\ref{fig:logit_difference}. Figure~\ref{fig:budget_logit_difference} shows the Log-Scaled LD of WB and Ours on different $\rho$. We found that a larger $\rho$ provides a better poisoning attacker. When $\rho$ is no smaller than 0.04, our method (solid lines) can successfully poison the AT. In comparison, the WB method hardly poisons AT even when $\rho$ is set to a large value (e.g., 0.08).

\noindent\textbf{Poisoning transferability across different network structures.}
We evaluate the transferability of the poisoned data across different network structures by counting the number of target points misclassified as the adversarial class by a robust learner over repeated trials.
The success rate is defined as the number of successful poisoning cases divided by the total number of trials.

Over the repeated trials, we count the number of target points misclassified as the adversarial class by a robust learner. 
The success rate refers to the number of successful poisoning cases over the total number of trials.  
We use the natural poisoning success rate on $x_{\tar}$ ($nat. ~success$) and the adversarial poisoning success rate on $\xadv_{\tar}$ ($adv.~success$) as evaluation metrics in Table~\ref{table:poisoning_success_rate}. 
This metric follows \cite{geiping_witches_2021}.
We use the natural poisoning success rate on $x_{\tar}$ (denoted as $nat.~success$) and the adversarial poisoning success rate on $\xadv_{\tar}$ (denoted as $adv.~success$) as evaluation metrics, following \cite{geiping_witches_2021}.

Table~\ref{table:poisoning_success_rate} reports the poisoning success rates of the robust learners using ResNet-18 and VGG-11, respectively. All poisoned data are generated by the attackers using the robust ResNet-18.
We observe that the poisoned data exhibits some transferability across different network structures.
For example, the poisoned data generated based on ResNet-18 has some effectiveness in VGG-11.
Compared to the natural target $x_{\tar}$, its adversarial variant $\xadv_{\tar}$ demonstrates better transferability.
%
%
\begin{table}[tp!] 
        \vspace{0mm}
	\centering \scriptsize
	\renewcommand\arraystretch{1}
	\caption{Comparison to WB with an $\ell_{\infty}$ bound of $\epsilon = 16/255$ and a poison budget $\rho$ of $0.04$ for CIFAR-10, $0.005$ for Tiny ImageNet. Note that \textit{nat. success} refers to poisoning success rate on the natural data, and \textit{adv. success} refers to the poisoning success rate on the adversarial data.}
	\label{table:poisoning_success_rate}
        \vspace{-2mm}
	\setlength{\tabcolsep}{0.7mm}{
		\begin{tabular}{cc|cc|cc}
			\toprule
			\multirow{2}*{Dataset} & \multirow{2}*{Base Network} & \multicolumn{2}{c}{WB} & \multicolumn{2}{c}{Ours} \\
			\cmidrule(lr){3-4}\cmidrule(lr){5-6} & &\textit{nat. success}&\textit{adv. success}&\textit{nat. success}&\textit{adv. success} \\
			\midrule
			\multirow{2}*{CIFAR-10} & ResNet-18 & 0.0\% & 2.5\% & 62.5\% & 62.5\% \\
			& VGG-11 & 0.0\% & 0.0\% & 0.0\% & 10.0\% \\
			\midrule
			\multirow{2}*{Tiny ImageNet} & ResNet-18 & 32.5\% & 35.0\% & 97.5\% & 97.5\% \\
			& VGG-11 & 0.0\% & 50.0\% & 0.0\% & 80.0\% \\
			\bottomrule 
	\end{tabular}}
\end{table}

\noindent\textbf{Evaluation of our poisoning method against ST.}
Although our work primarily focuses on poisoning the AT, we are evaluate whether our poisoning method can be effective against the ST. 
We compare our method with WB, which is designed specifically for poisoning the ST.
We compare with the WB that focuses on poisoning the ST. Table~\ref{table:st_poisoning_success_rate} reports the poisoning success rate of WB and Ours on CIFAR-10.
Table~\ref{table:st_poisoning_success_rate} reports the poisoning success rates of WB and our method on CIFAR-10.
\begin{table}[tp!]
        \vspace{-2mm}
	\centering \footnotesize
	\caption{Poisoning with an $\ell_\infty$ bound of $\epsilon = 16/255$ and a poison budget $\rho = 0.04$.}
	\label{table:st_poisoning_success_rate}
        \vspace{-2mm}
	\setlength{\tabcolsep}{5mm}{
		\begin{tabular}{cc|c|c}
			\toprule
			Dataset & Base Network & WB & Ours \\
			\midrule
			CIFAR-10 & ResNet-18 & 95.0\% & 40.0\% \\
			\bottomrule 
	\end{tabular}}
        \vspace{-2mm}
\end{table}
We find that our poisoned data generated using the robust model can still effectively poison the ST, but not as strongly as WB. This is consistent with prior research that suggests AT relies more on robust features that contain semantic meaning, while ST relies more on non-robust features that are visually similar to random noise~\cite{tsipras19_robustness_at_odd}.
As shown in Figure~\ref{fig:samples}, our poisoned noises contain some semantic meaning, while the WB noises are more like random noise.
Given that AT and ST rely on different features for prediction, poisoning strategies should be adapted accordingly.


\begin{figure}[h!]
        \vspace{-2mm}
	\centering
	\includegraphics[scale=0.24]{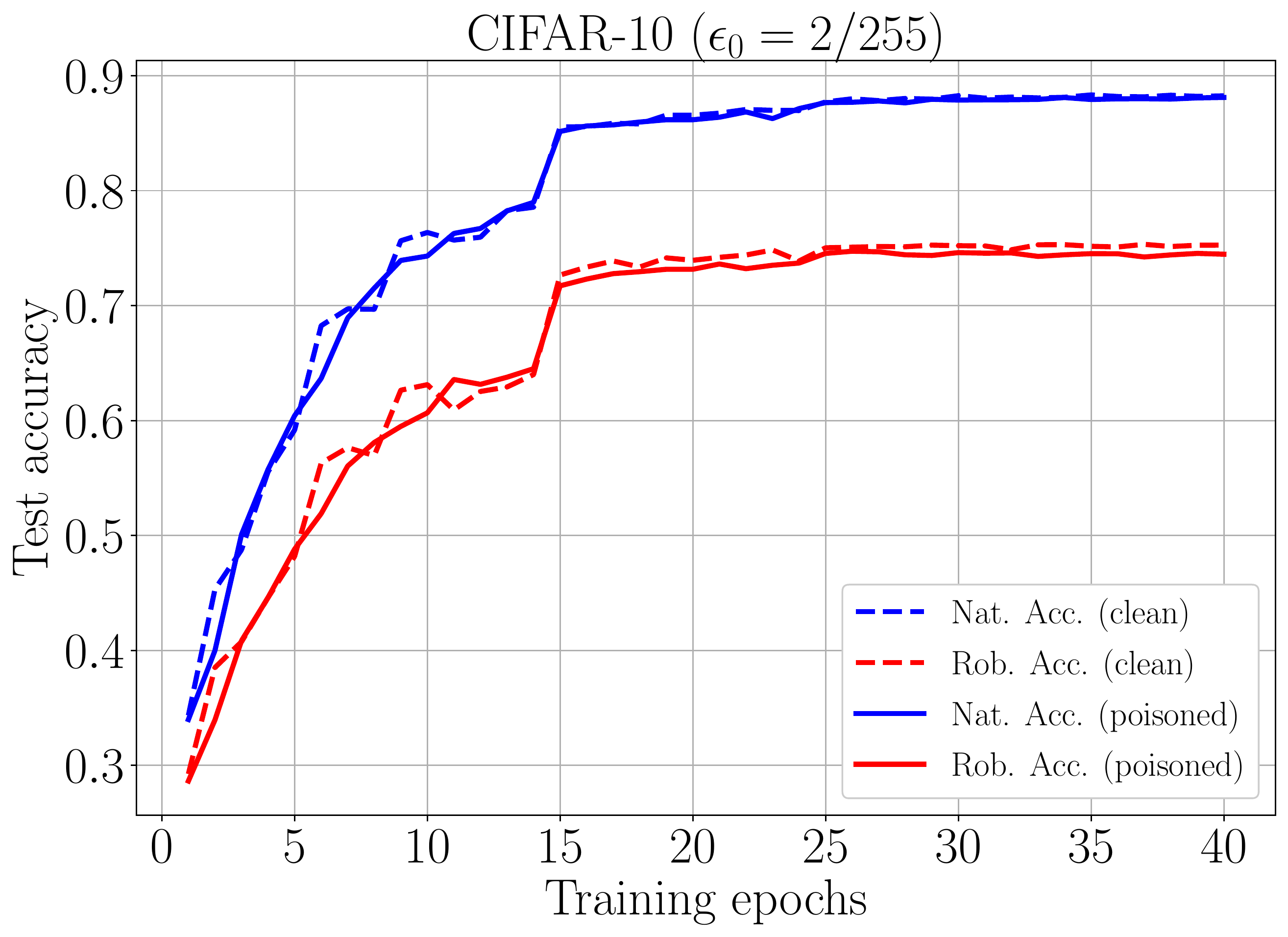} \\
	\vspace{-2mm}
	\caption{Test accuracy curves of robust learner $\A_{\epsilon_{0}=2/255}$ on clean training set $S$ and poisoned training set $S'$. Nat. Acc. refers to natural test accuracy on all natural test data, and Rob. Acc. refers to robust test accuracy on the adversarial test data.}
	\label{fig:tar_valid_learning_curve}
\end{figure}
\noindent\textbf{The poisons are hard to detect.}
This part highlights the insidious nature of our targeted attacks, which aim to control the behavior of a specific target data without degrading the overall classification performance, making it hard to detect.
We demonstrate the test accuracy curves of the robust learner $\A_{\epsilon_{0}=2/255}$ on the clean training set $S$ and the poisoned training set $S'$ in Figure~\ref{fig:tar_valid_learning_curve}. Here, $S'$ is generated based on a perturbation radius $\epsilon = 16/255$ and poison budget $\rho = 0.04$. The validation curves of the poisoned set $S'$ and clean set $S$ are almost indistinguishable, confirming that our poisons are hard to detect even with the validation set.
\begin{figure}[h!]
        \vspace{-0mm}
	\centering
	\includegraphics[scale=0.24]{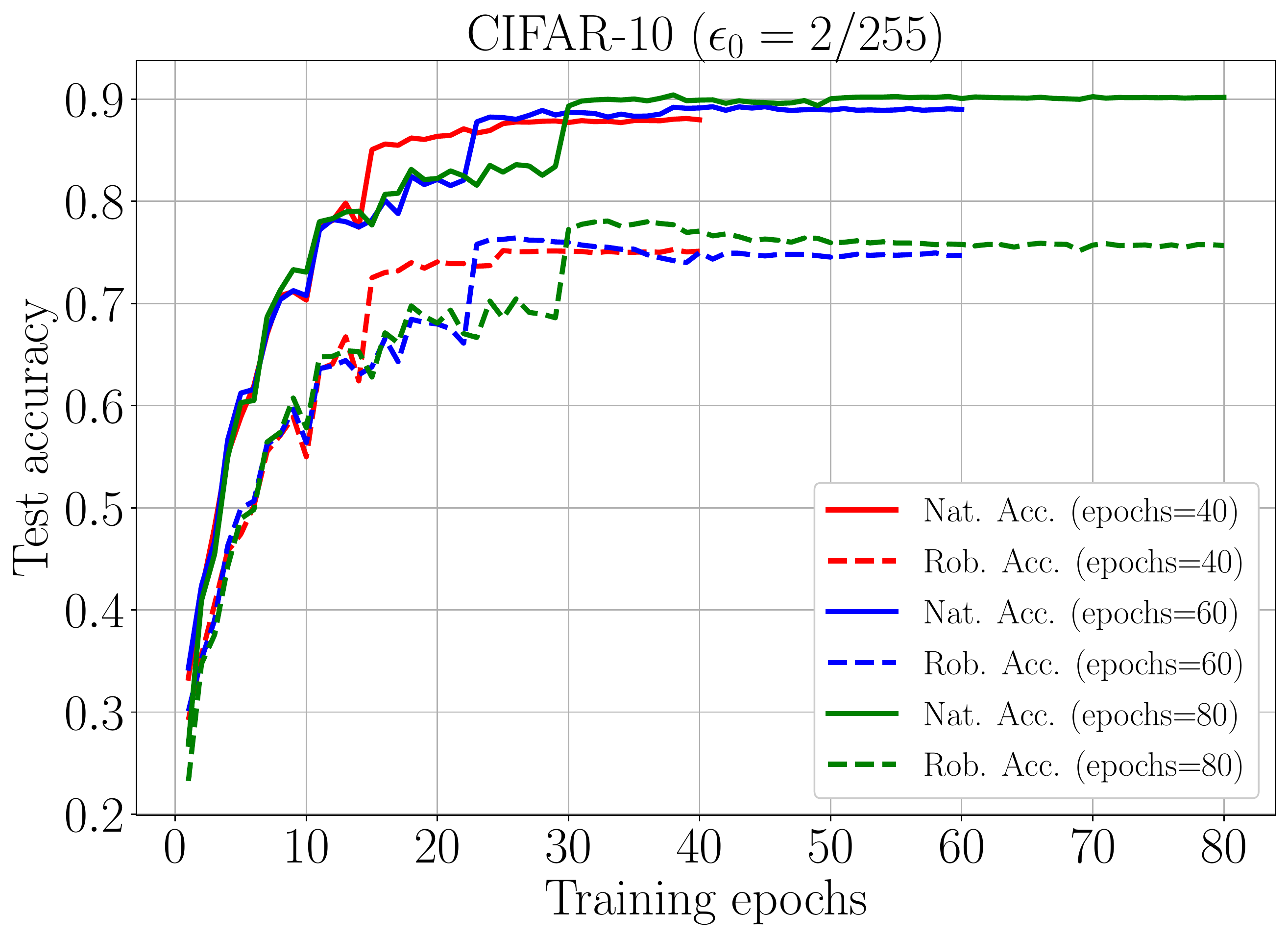} \\
	\vspace{-2mm}
	\caption{Test accuracy curves of robust learner $\A_{\epsilon_{0}=2/255}$ with different numbers of training epochs on the same poisoned set $S'$.}
	\label{fig:tar_valid_learning_curve_overfit}
        \vspace{-2mm}
\end{figure}
\begin{table}[h!]
        \vspace{-4mm}
	\centering \scriptsize
	\renewcommand\arraystretch{1}
	\caption{Poisoning success rate under different numbers of training epochs.}
	\label{table:overfit_poisoning_success_rate}
        \vspace{-2mm}
	\setlength{\tabcolsep}{2.6mm}{
		\begin{tabular}{ccc|cc}
			\toprule
			\multirow{2}*{Dataset} & \multirow{2}*{Base Network} & \multirow{2}*{Epochs} & \multicolumn{2}{c}{Ours} \\
			\cmidrule{4-5} & & &\textit{nat. success}&\textit{adv. success} \\
			\midrule
			\multirow{3}*{CIFAR-10} & \multirow{3}*{ResNet-18} & 40 & 0.0\% & 0.0\% \\
			& & 60 & 62.5\% & 100.0\% \\
			& & 80 & 100.0\% & 100.0\% \\
			\bottomrule 
	\end{tabular}}
        \vspace{-4mm}
\end{table}

\noindent\textbf{Robust overfitting amplifies the toxicity of the poisons}
We observed that the same poisoned data can have a stronger toxicity on robust learners when robust overfitting~\cite{rice2020overfitting} occurs.
In this section, we evaluate robust learner $\A_{\epsilon_{0}=2/255}$ on the same poisoned set $S'$ with different numbers of training epochs. $S'$ is generated with $\epsilon = 16/255$ and $\rho = 0.01$, corresponding to the random seed 2110000000. We perform AT eight times for each evaluation. 

Figure~\ref{fig:tar_valid_learning_curve_overfit} shows the test accuracy curves of $\A_{\epsilon_{0}=2/255}$ with different numbers of training epochs. We observe that the robust test accuracy (blue or green dashed lines) gradually decreases after reaching a peak when the number of epochs is 60 or 80, indicating the occurrence of robust overfitting in the training process. However, robust overfitting does not occur when the number of epochs is 40.

Table~\ref{table:overfit_poisoning_success_rate} reports the poisoning success rate of $S'$ under different numbers of training epochs. We find that it is difficult for $S'$ to fool $\A_{\epsilon_{0}=2/255}$ without robust overfitting. However, when robust overfitting occurs, the behavior of $\A_{\epsilon_{0}=2/255}$ can be easily controlled by $S'$.

\noindent\textbf{Evaluating the effectiveness of our targeted attack on different AT strategies.}
In this section, we evaluate whether our attack is capable of poisoning various AT strategies. 
We first generate a poisoned set using an attacker trained by standard AT, and then apply friendly adversarial training (FAT)~\cite{zhang2020fat} and fast adversarial training (FastAT)~\cite{wong2020fast_zico_kolter} on $S'$. 
Both FAT and FastAT are based on ResNet18 using SGD with a momentum of 0.9 and a weight decay of 0.0005, trained for 40 epochs. 
For FAT, we use a learning rate of 0.1 and decay by a factor of 0.1 at the 15th and 30th epochs. For FastAT, we use a cyclic learning rate scheduler with a maximum learning rate of 0.2. We set the step size of FAT to $\epsilon_{0}/4$, and FastAT to $\epsilon_{0}$. We evaluate the attack by conducting an AT 3 times each at ten different poison settings, and report the poisoning success rate in Table~\ref{table:at_transfer_poisoning_success_rate}. 
The results demonstrate the efficacy of our attack against robust learners trained using different AT strategies.
\begin{table}[h!]
        \vspace{-0mm}
	\centering \scriptsize
	\renewcommand\arraystretch{1}
	\caption{Poisoning success rate on robust learner $\A_{\epsilon_{0}=2/255}$ trained with different AT strategies.}
	\label{table:at_transfer_poisoning_success_rate}
        \vspace{-2mm}
	\setlength{\tabcolsep}{2.4mm}{
		\begin{tabular}{ccc|cc}
			\toprule
			\multirow{2}*{Dataset} & \multirow{2}*{Base Network} & \multirow{2}*{AT Strategy} & \multicolumn{2}{c}{Ours} \\
			\cmidrule{4-5} & & &\textit{nat. success}&\textit{adv. success} \\
			\midrule
			\multirow{2}*{CIFAR-10} & \multirow{2}*{ResNet-18} & FAT & 56.7\% & 40.0\% \\
			& & FastAT & 56.7\% & 36.7\% \\
			\bottomrule 
	\end{tabular}}
        \vspace{-4mm}
\end{table}

\subsection{Clean-label Untargeted Attacks Against AT}
\label{sec_exp:clean-label-untar}
\begin{figure*}[t!]
	\vspace{0mm}
	\centering
	\subfigure[CIFAR-10]{
		\begin{minipage}[t]{0.48\linewidth}
			\centering
			\includegraphics[scale=0.7]{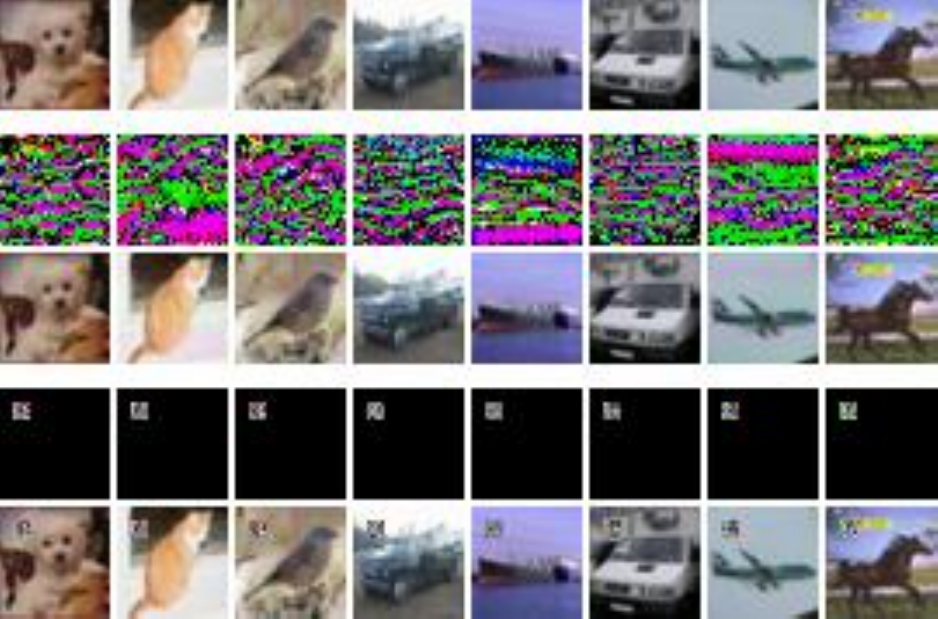}
		\end{minipage}
	}
	\subfigure[CIFAR-100]{
		\begin{minipage}[t]{0.48\linewidth}
			\centering
			\includegraphics[scale=0.7]{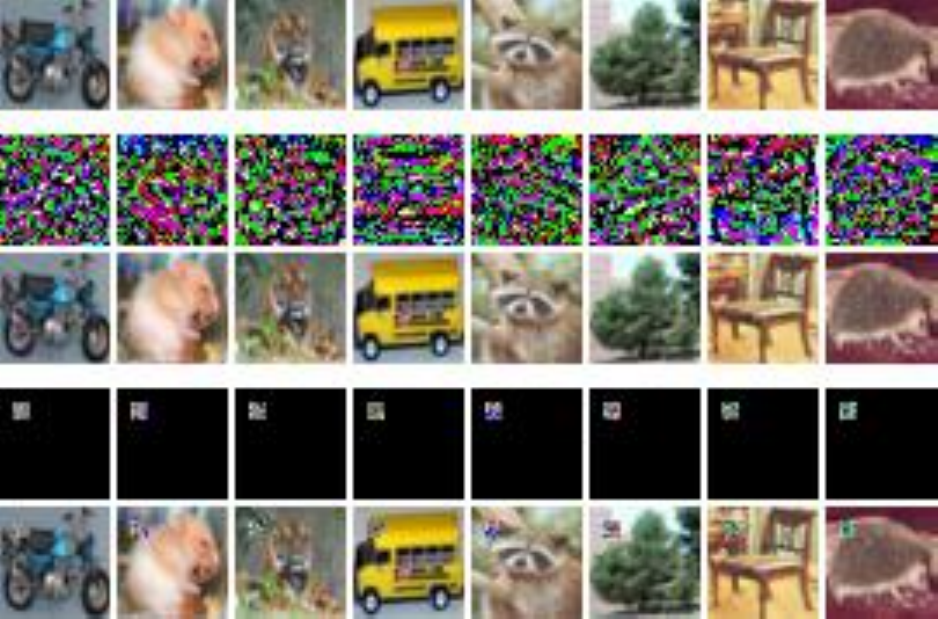}
		\end{minipage}
	}
	\vspace{-2mm}
	\caption{Comparison of clean-label untargeted poisoning attacks generated by robust error-minimizing noise (REM) and our method (sticker by Algorithm~\ref{alg:clean-label-untar}). The top row displays the original natural images. The second and third rows show the imperceptible noise (scaled by 31 times for visualization) and the corresponding poisoned images generated by REM. The fourth and fifth rows display the human-visible stickers and our poisoned images.}
	\label{fig:samples_patch}
        \vspace{-2mm}
\end{figure*}

\begin{table*}[h!]
	\small
	\centering
	\renewcommand\arraystretch{1}
	\caption{The best natural/robust test accuracy (\%) ($\pm$ standard deviations) of the robust learners $\A_{\epsilon_{0}}$ on clean set $S$ and poisoned set $S'$ by REM and Algorithm~\ref{alg:clean-label-untar} (Ours), respectively. The complete learning curves over training epochs are in the appendix.
	}
	\label{exp:un-tar-attack_main-result}
        \vspace{-2mm}
	\setlength{\tabcolsep}{2mm}{
		\begin{tabular}{cc|cc|cc|cc}
			\toprule
			\multirow{2}*{Dataset} & \multirow{2}*{$\epsilon_{0}$} &\multicolumn{2}{c}{Clean}&\multicolumn{2}{c}{REM}&\multicolumn{2}{c}{Ours} \\	
			\cmidrule(lr){3-4}\cmidrule(lr){5-6}\cmidrule(lr){7-8}
			&&Nat.&Rob.&Nat.&Rob.&Nat.&Rob. \\
			\midrule
			\multirow{4}*{CIFAR-10}&$4/255$&88.78&67.22&51.42$\pm$1.36(\textcolor{teal}{-37.36})&31.93$\pm$0.85(\textcolor{teal}{-35.29})&70.30$\pm$0.62(\textcolor{teal}{-18.48})&50.58$\pm$0.64(\textcolor{teal}{-16.64})\\
			&$8/255$&81.45&50.47&83.68$\pm$0.19(\textcolor{red}{+2.23})&37.92$\pm$0.31(\textcolor{teal}{-12.55})&63.69$\pm$0.63(\textcolor{teal}{-17.75})&38.39$\pm$0.39(\textcolor{teal}{-12.08}) \\
			&$12/255$&72.26&40.03&74.08$\pm$0.64(\textcolor{red}{+1.18})&36.97$\pm$0.16(\textcolor{teal}{-3.06})&66.02$\pm$1.49(\textcolor{teal}{-6.24})&36.61$\pm$0.72(\textcolor{teal}{-3.41}) \\
			&$16/255$&62.01&32.68&63.58$\pm$0.58(\textcolor{red}{+1.57})&31.82$\pm$0.28(\textcolor{teal}{-0.85})&60.40$\pm$0.61(\textcolor{teal}{-1.61})&32.04$\pm$0.51(\textcolor{teal}{-0.64}) \\
			\midrule
			\multirow{4}*{CIFAR-100}&$4/255$&64.43&39.20&40.68$\pm$0.53(\textcolor{teal}{-23.75})&22.30$\pm$0.77(\textcolor{teal}{-16.90})&39.69$\pm$0.67(\textcolor{teal}{-24.74})&25.93$\pm$0.34(\textcolor{teal}{-13.27})\\
			&$8/255$&56.36&27.95&56.74$\pm$0.14(\textcolor{red}{+0.38})&26.55$\pm$0.05(\textcolor{teal}{-1.40})&51.11$\pm$2.40(\textcolor{teal}{-5.24})&25.82$\pm$0.73(\textcolor{teal}{-2.13}) \\
			&$12/255$&47.44&21.20&48.80$\pm$0.16(\textcolor{red}{+1.36})&20.44$\pm$0.10(\textcolor{teal}{-0.75})&47.11$\pm$0.31(\textcolor{teal}{-0.32})&20.80$\pm$0.11(\textcolor{teal}{-0.39}) \\
			&$16/255$&38.64&17.17&39.73$\pm$0.36(\textcolor{red}{+1.08})&16.51$\pm$0.10(\textcolor{teal}{-0.65})&37.95$\pm$0.12(\textcolor{teal}{-0.69})&16.94$\pm$0.13(\textcolor{teal}{-0.22}) \\
			\bottomrule 
	\end{tabular}}
        \vspace{-2mm}
\end{table*}

\textbf{Data augmentation.}
We conduct experiments on CIFAR-10, CIFAR-100 and subset of ImageNet. 
To ensure fair comparisons with REM~\cite{fu2022robust}, we apply random crop, random flip, and rescaling of each pixel to $[-0.5, 0.5]$ on each training data when the robust learner performed AT. 
For training the generator, we apply rescaling per pixel to $[-0.5, 0.5]$ on each training data without using multiple augmentation techniques as extensively used in REM.

\noindent\textbf{Robust learner.}
The robust learners employ standard AT. We apply the $\ell_{\infty}$ the PGD method to generate adversarial examples for both the training and the testing. The PGD method has random initialization enabled. 
Specifically, for CIFAR-10 and CIFAR-100, we use 10 steps and a step size of $\epsilon_{0}/5$, while for ImageNet subset, we use 8 steps and a step size of $\epsilon_{0}/4$.
All models are trained with a batch size of 128, a weight decay factor of 0.0005, and SGD with a momentum of 0.9 and an initial learning rate of 0.1. For CIFAR-10 and CIFAR-100, the models are trained for 15000 iterations with a learning rate scheduler that reduces the learning rate by a factor of 0.1 every 6000 iterations. For ImageNet Subset, the models are trained for 40000 iterations with a learning rate scheduler that reduces the learning rate by a factor of 0.1 every 16000 iterations.


\noindent\textbf{The REM implementation details.}
We implement the baseline REM method~\cite{fu2022robust} by training a noise generator $g_{\epsball}$ according to Eq.~\ref{eq:train_gen} with $\epsilon_{0} = 4/255$ and $\epsilon = 8/255$ as in the original paper.
We use ResNet-18 to train the generator for 5000 iterations. We also use the \textit{expectation over transformation} technique, which involves repeated sampling of augmented samples, and set the number of repeated samplings to 5 to keep it consistent with the original paper.
%
Finally, we use the generator to obtain the robust error-minimizing noise that is bounded by $\ell_{\infty}$-norm with the size of $\epsilon = 8/255$.

\noindent\textbf{Generation of stickers.}
We train the sticker generator on both CIFAR-10 and CIFAR-100 datasets using a ResNet-18 model. The generator is trained for 5000 iterations using SGD with 0.9 momentum and an initial learning rate of 0.1, where the learning rate decays by a factor of 0.1 every 2000 iterations. The "Patch" is initialized as a square with a size of $3\%$ of the original image. We update the "Patch" using the PGD method with a step size of $35/255$ and perturbation steps of $10$. We place the stickers on the upper left area of each image. The generated stickers are visually small and do not obscure the semantic meanings, as shown in Figure~\ref{fig:samples_patch}.

\noindent\textbf{Evaluation details.}
We evaluate the performance of stickers (using $\ell_{0}$-norm ball that changes $3\%$ of overall pixels) and compare it with REM (using $\ell_{\infty}$-norm ball and $\epsilon=8/255 \approx{0.031}$) on robust learners with $\epsilon_{0} \in [4/255, 16/255]$. For each $\epsilon_{0}$, we execute $\epsballzero$-AT five times with different random seeds. To ensure a consistent result, we fix the random seeds 2000000000-2111100000.

Throughout the training epochs, we record the best natural test accuracy on all test data $x$ and the best robust test accuracy on adversarial data $\xadv \in \epsballzero[x]$. 
Rather than relying on the last-checkpoint accuracy, we chose to do this because we found that the AT has a catastrophic overfitting~\cite{wong2020fast_zico_kolter} when learning from the poisoned set $S'$ (see appendix), which gives a false sense of the effectiveness of the poisoning attacks. This could be easily avoided by the robust learner using early stopping~\cite{rice2020overfitting} on the validation set. In the appendix (see Figure~\ref{fig:overfitting}), we report the full learning curves of CIFAR-10 experiments. 
Additionally, we evaluated our stickers on the ImageNet Subset dataset (see below).

\subsubsection{Main Results of Untargeted Poisoning Attacks}
\label{sec:untar_main_results}
\noindent\textbf{CIFAR-10 and -100 results.}
Table~\ref{exp:un-tar-attack_main-result} shows that as $\epsilon_{0} \to \epsilon$ (e.g., $\epsilon_{0}= \epsilon = 8/255$), the effectiveness of REM, which uses the same norm ball $\epsball$ as the robust learner $\epsballzero$, decreases in generating poisoned data to deter (highlighted by ``$-$'' in cyan-green) unauthorized data collection, and sometimes even has an encouragement effect (highlighted by ``$+$'' in red). 
In contrast, our method using a different $\epsball'$ remains effective in deterrence when $\epsilon_{0}$ approaches $\epsilon$, which supports the efficacy of specifying a norm ball different from the learner's.
Note that as $\epsilon_{0}$ becomes larger than $\epsilon$, both REM and our method have diminishing returns, probably due to strong invariance~\cite{tramer2020fundamental} and low natural and robust accuracies under large perturbation radii, which calls for large radius $\epsilon$ poisons.
Complete results, including learning curves over epochs, are provided in the appendix (see Figure~\ref{fig:overfitting}).

\begin{figure*}[tp!]
	\centering
	\includegraphics[scale=0.215]{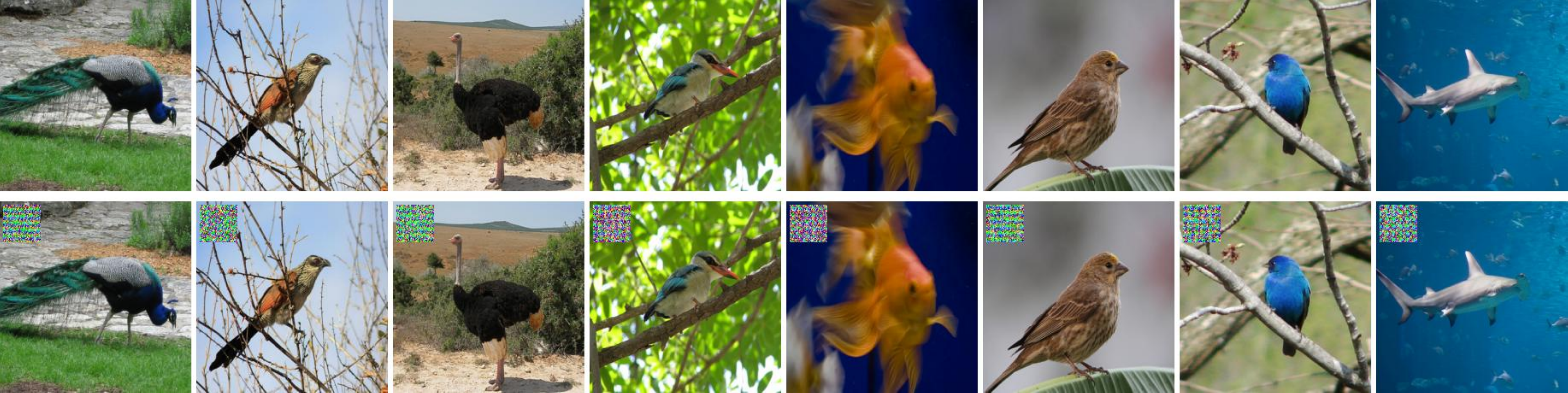} \\
	\vspace{-2mm}
	\caption{Visualization of clean-label untargeted poisoning attacks by our sticker on ImageNet Subset. The first row has unperturbed natural images. The second row has our poisoned images attached with stickers.}
	\label{fig:imgnet_samples_patch}
        \vspace{-4mm}
\end{figure*}
\begin{figure}[h]
	\centering
	\includegraphics[scale=0.24]{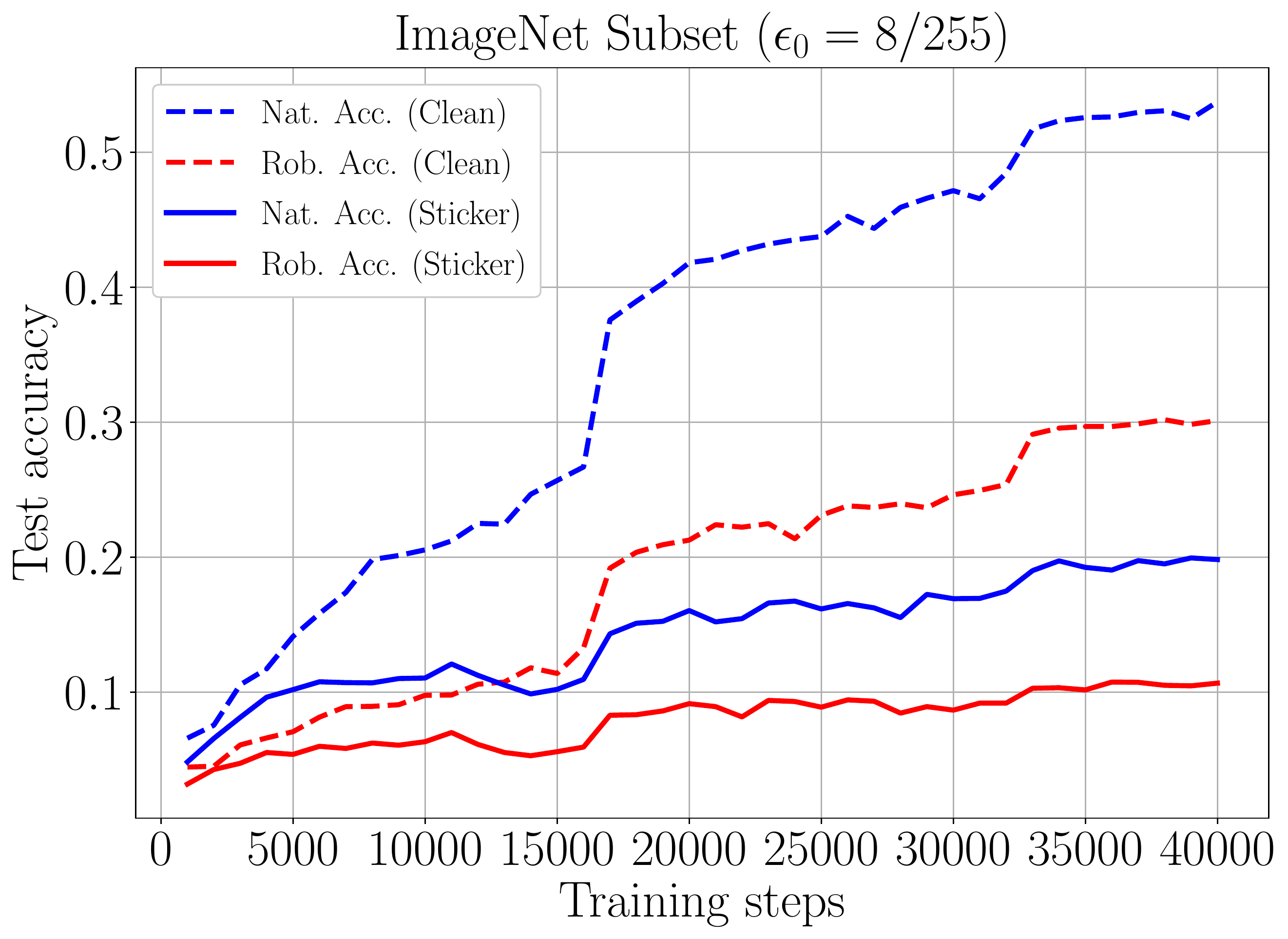} \\
	\vspace{-2mm}
	\caption{Test accuracy curves of robust learner $\A_{\epsilon_{0}=8/255}$ on the clean set $S$ and the poisoned set $S'$.}
	\label{fig:un_tar_imgnet_learning_curve_8}
        \vspace{-2mm}
\end{figure}

\noindent\textbf{ImageNet subset results.}
We visualize and evaluate the performance of our sticker on the ImageNet subset in Figures~\ref{fig:imgnet_samples_patch} and~\ref{fig:un_tar_imgnet_learning_curve_8}.
The ImageNet Subset contains the first 100 classes of the full ImageNet dataset. We use SGD with 0.9 momentum and an initial learning rate of 0.1 to train the sticker generator. The generator undergoes 3000 iterations, and we decay the learning rate by a factor of 0.1 every 1200 iterations. We initialize the "Patch" as a square with a size of $3\%$ of the original image and update it using the PGD method with a step size of $35/255$ and $7$ iterations. Finally, we evaluate the performance of the stickers on the robust learner $\A_{\epsilon_{0}=8/255}$.

The Figure~\ref{fig:un_tar_imgnet_learning_curve_8} presents the test accuracy curves of $\A_{\epsilon_{0}=8/255}$ on the clean set $S$ and the poisoned set $S'$ (modified by our stickers). The test accuracy curves on $S'$ (solid line) are significantly lower than those on $S$ (dashed line), which confirms the effectiveness of our stickers in degrading the robust learner. Additionally, we provide visualizations of the clean and poisoned data of the ImageNet Subset in Figure~\ref{fig:imgnet_samples_patch}. 
On the high-quality images, our small stickers do not obfuscate the semantic meaning, but are still effective in degrading overall performance of robust learners, which is a strong signal for preventing unauthorized data collection

\subsubsection{Ablation Studies}
\begin{figure}[h!]
        \vspace{-2mm}
	\centering
	\includegraphics[scale=0.24]{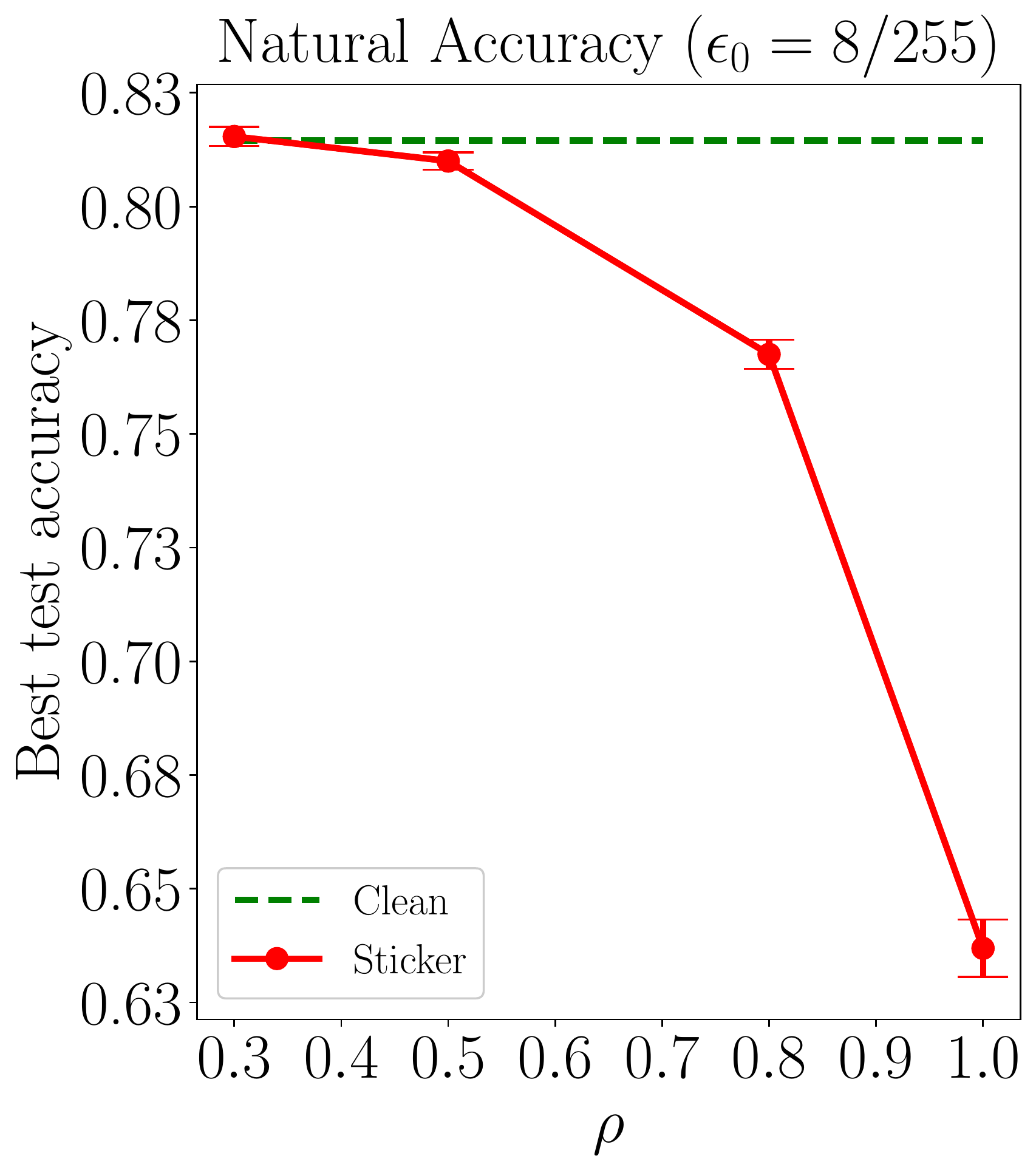}
	\hspace{2mm}
	\includegraphics[scale=0.24]{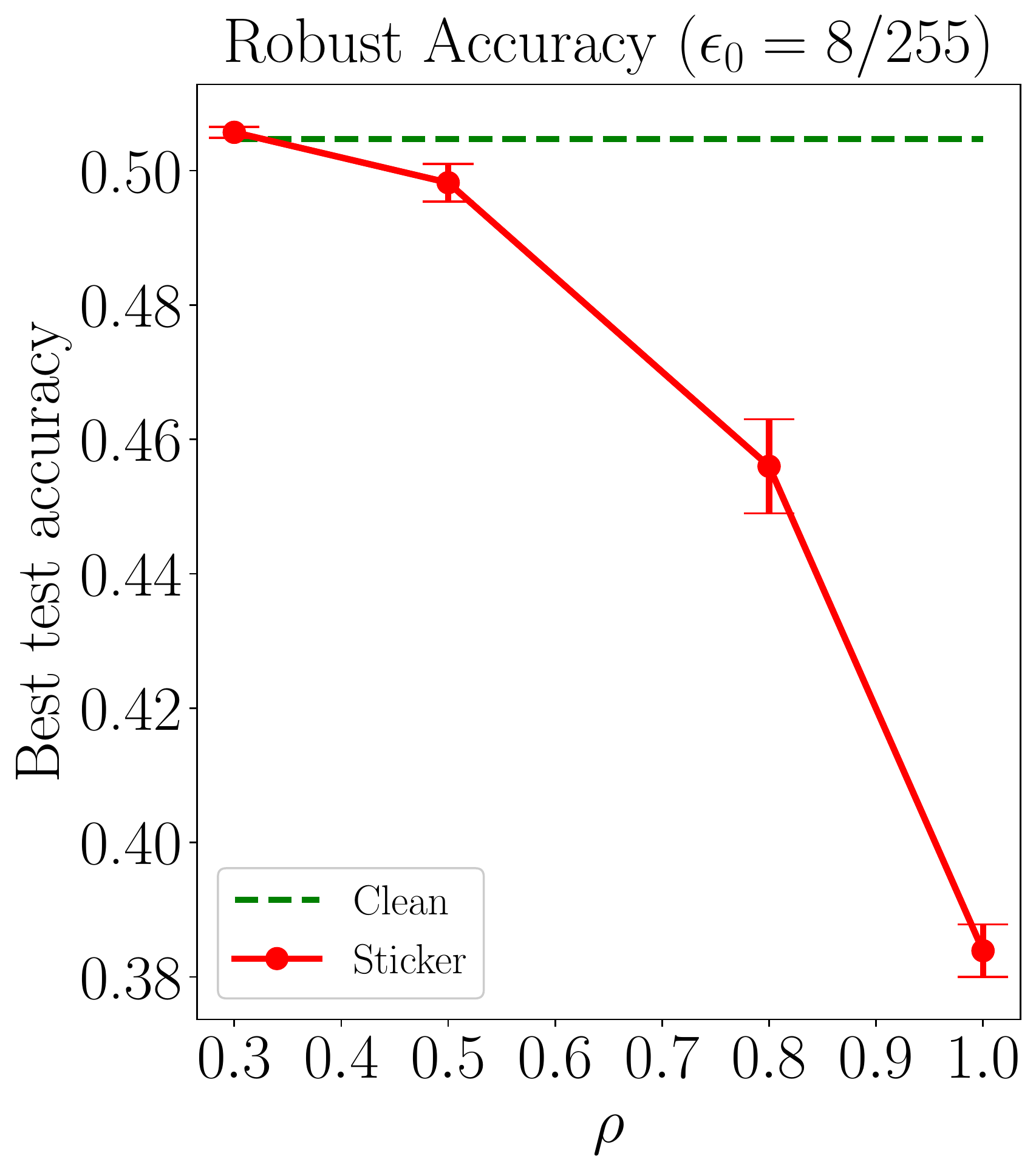} \\
	\vspace{-2mm}
	\caption{Best natural test accuracy on Privacy sticker with various $\rho$ on CIFAR-10 dataset. Green dashed lines are the oracle that indicates the best natural test accuracy of $\A_{\epsilon_{0}=8/255}$ on the clean set $S$.}
	\label{fig:budget_best_accuracy}
    \vspace{-2mm}
\end{figure}

\noindent\textbf{Different poison budget.}
In Section~\ref{sec:untar_main_results}, we only focus on poisoning the entire training set. 
In this section, we consider a more challenging scenario. We consider a smaller poison budget $\rho$. We firstly randomly choose $\rho \times |S|$ data to learn a sticker generator and then use the constructed generator to attach stickers on the $\rho$ portion data. Then, we replace the clean $\rho$ portion with the poisoned $\rho$ portion to construct a poisoned set $S'$ (Note that $|S'| = |S|$). 
Then, the robust learner performs AT, and we only collect the best test accuracy. We repeat robust learner five times and report the median and the standard deviation plotted as an error bar, as shown in Figure~\ref{fig:budget_best_accuracy}. 
From Figure~\ref{fig:budget_best_accuracy}, we found it is not necessary to add stickers to all training data, and the robust learner's performance can be degraded even when a portion of the training data has stickers.

\noindent\textbf{Time costs of training generators.}
We calculate the time of training the generators for REM and sticker, respectively. We report the results in Table~\ref{table:time_cost}. Compared with REM, training our sticker generator is much more efficient. 
\begin{table}[htb]
        \vspace{-3mm}
	\centering
	\renewcommand\arraystretch{1.2}
	\caption{Computational costs of training the generators for REM and sticker, respectively.}
	\label{table:time_cost}
        \vspace{-2mm}
	\setlength{\tabcolsep}{4mm}{
		\begin{tabular}{c|cc}
			\toprule
			Dataset & REM & Sticker \\
			\midrule
			CIFAR-10 / 100 & 32.4h & 0.5h \\
			\bottomrule 
	\end{tabular}}
        \vspace{-1mm}
\end{table}

\begin{figure}[h!]
	\centering
	\includegraphics[scale=0.24]{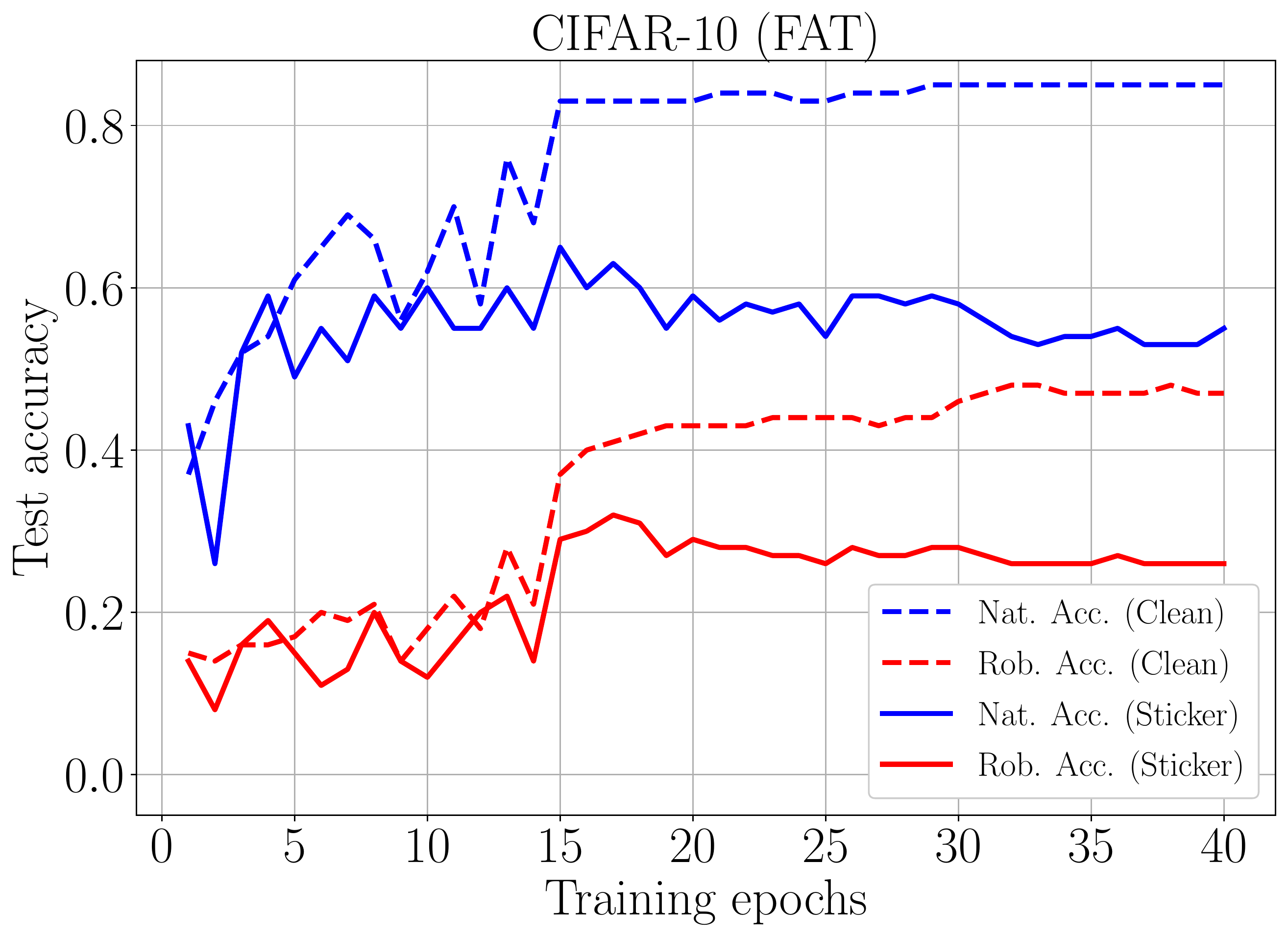}
	\\
	\includegraphics[scale=0.24]{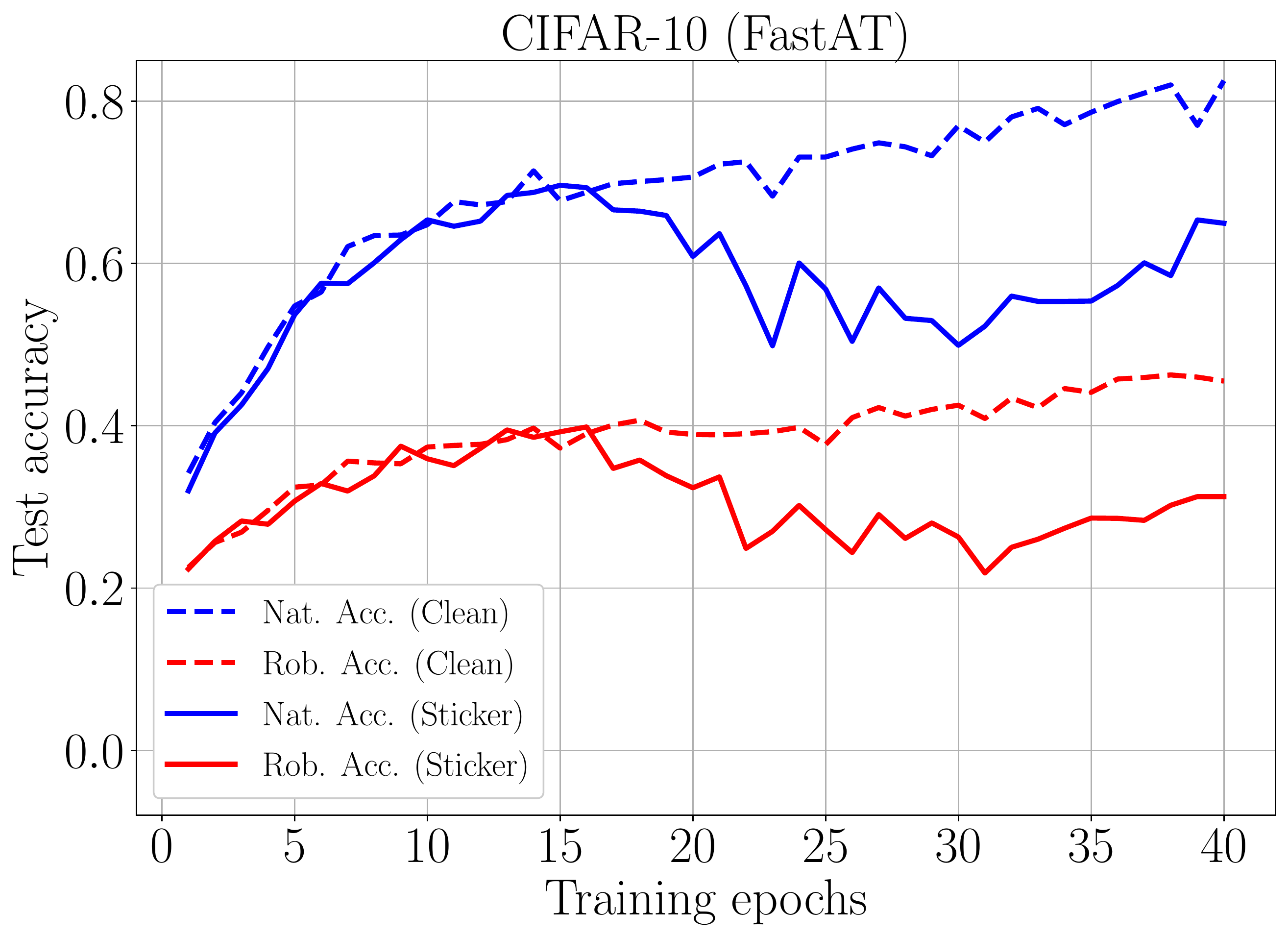} \\
	\vspace{-2mm}
	\caption{Test accuracy curves of robust learners trained with different AT strategies.}
	\label{fig:untar_diff_at}
        \vspace{-4mm}
\end{figure}

\noindent\textbf{Evaluation on different AT strategies.}
In this section, we use robust learners trained with different AT strategies to evaluate the effectiveness of stickers. We run FAT and FastAT on clean set $S$ and poisoned set $S'$, respectively. Both FAT and FastAT are based on ResNet18 using SGD with a momentum of 0.9 and a weight decay of 0.0005 to train 40 epochs, and the perturbation radius is $8/255$. For FAT, we use a learning rate of 0.1 and decay by a factor of 0.1 at the 15th and 30th epochs. For FastAT, we use a cyclic learning rate scheduler with a maximum learning rate of 0.2. Figure~\ref{fig:untar_diff_at} shows the test accuracy curves of different AT strategies on CIFAR-10. The test accuracy on $S'$ (solid line) is lower than the test accuracy on $S$ (dashed line), whether it is natural test accuracy or robust test accuracy. This indicates that the sticker has good transferability among robust learners using different AT strategies.

\section{Conclusion}
\label{sec:conclusion}
This paper has provided insights into the vulnerabilities of adversarial training (AT), which challenges the notion that AT is an effective defense against imperceptible noises. While this work highlights the potential negative impacts of poisoning attacks on machine learning systems, it also sheds light on the importance of understanding the system's vulnerabilities to design more reliable defenses. We believe that our findings can contribute to the development of more robust AT methods for high-stakes machine learning applications. Further research can explore the vulnerabilities of other types of AT and consider different attacker capabilities, such as label modification, to design corresponding defenses. Ultimately, the ongoing battle between attackers and defenders underscores the need for continuous efforts to improve the security of machine learning systems.

\bibliographystyle{IEEEtran}
\bibliography{reference_short}
%

%
\begin{IEEEbiography}[{\includegraphics[scale=0.16]{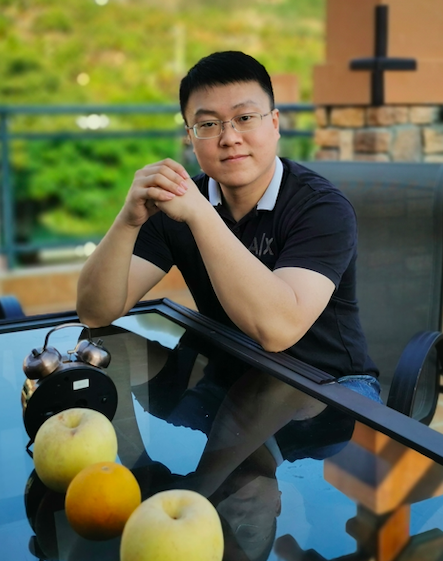}}] {Jingfeng Zhang} is a research scientist at the RIKEN Center for Advanced Intelligence Project, where he focuses on making artificial intelligence secure for human beings. He received his Ph.D. in computer science from the National University of Singapore in 2020 and his Bachelor's degree in computer science from Taishan College at Shandong University, China, in 2016. Jingfeng has published 15 papers in prestigious conferences and journals, such as ICML, ICLR, NeurIPS, and IEEE TDSC. He also serves as an associate editor of IEEE Artificial Intelligence and is actively involved in reviewing papers in diverse domains. With his expertise and dedication, Jingfeng is dedicated to the advancement of AI and its safe integration into our lives.
\end{IEEEbiography}
\begin{IEEEbiography}[{\includegraphics[scale=0.038]{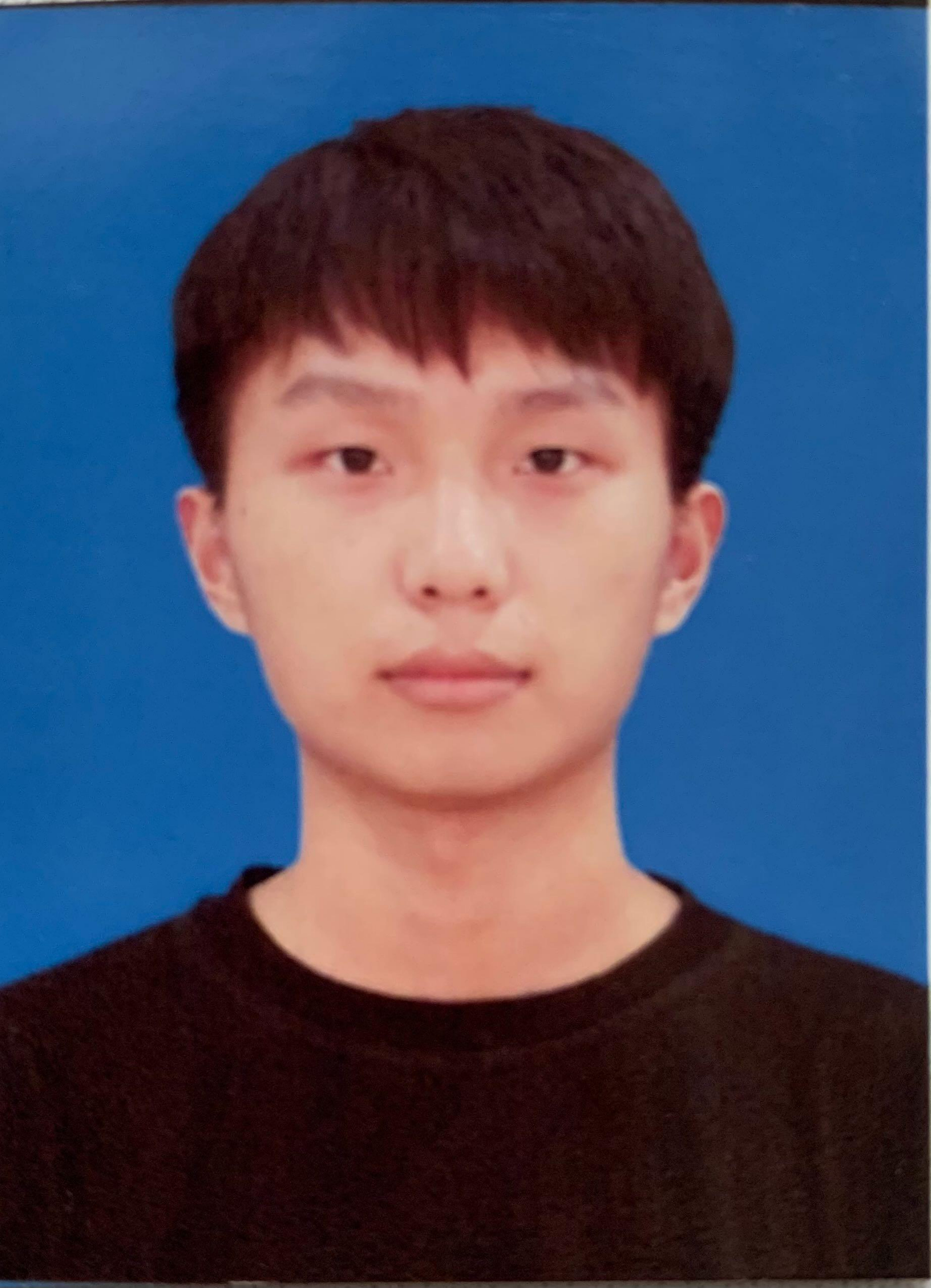}}]{Bo Song}
is currently a postgraduate student in the school of software, Shandong University, China. He received the BE degree in software engineering from Shandong University in 2021. His research interests include adversarial attacks, noisy-label learning.
\end{IEEEbiography}
\begin{IEEEbiography}[{\includegraphics[scale=0.33]{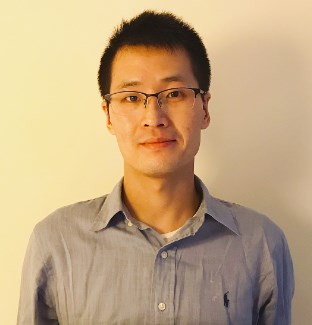}}]{Bo Han} is currently an Assistant Professor in Machine Learning and a Director of Trustworthy Machine Learning and Reasoning Group at Hong Kong Baptist University, and a BAIHO Visiting Scientist at RIKEN Center for Advanced Intelligence Project (RIKEN AIP). He was a Visiting Faculty Researcher at Microsoft Research (2022) and a Postdoc Fellow at RIKEN AIP (2019-2020). He received his Ph.D. degree in Computer Science from University of Technology Sydney (2015-2019). During 2018-2019, he was a Research Intern with the AI Residency Program at RIKEN AIP. He has co-authored a machine learning monograph, including Machine Learning with Noisy Labels (MIT Press). He has served as area chairs of NeurIPS, ICML, ICLR and UAI, and senior program committees of KDD, AAAI and IJCAI. He has also served as action (associate) editors of Transactions on Machine Learning Research and IEEE Transactions on Neural Networks and Learning Systems, and editorial board members of Journal of Machine Learning Research and Machine Learning Journal.
\end{IEEEbiography}

\begin{IEEEbiography}[{\includegraphics[scale=0.27]{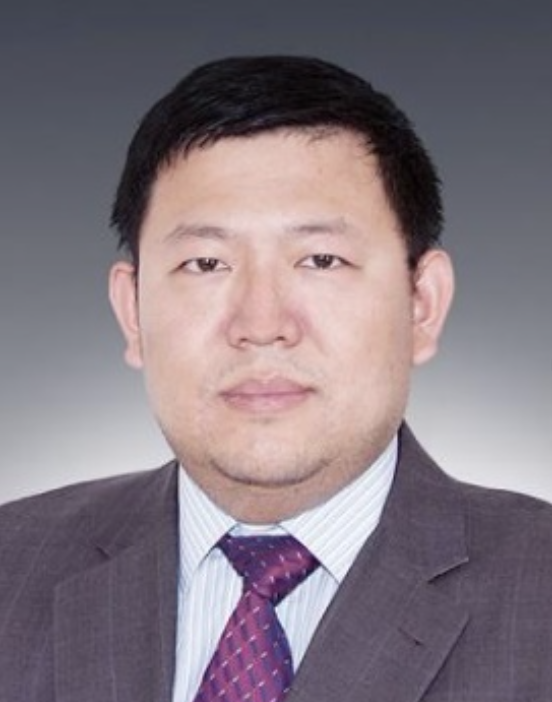}}]{Lei Liu} is a full professor in the school of software, Shandong University. He obtained the master and Ph.D degree in 2005 and 2010 from Bradford University, UK, respectively. Dr. LIU has published over 70 research papers on international conferences and journals. His research interest includes AI enabled network engineering, 5g technology, quality of service, AIoT.
\end{IEEEbiography}

\begin{IEEEbiography}[{\includegraphics[scale=0.45]{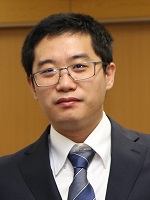}}]{Gang Niu} is currently an indefinite-term research scientist at RIKEN Center for Advanced Intelligence Project. He received the PhD degree in computer science from Tokyo Institute of Technology in 2013. Before joining RIKEN as a research scientist, he was a senior software engineer at Baidu and then an assistant professor at the University of Tokyo. He has published more than 90 journal articles and conference papers, including 31 ICML, 20 NeurIPS (1 oral and 3 spotlights), and 12 ICLR (1 outstanding paper honorable mention, 3 orals, and 1 spotlight) papers. He has co-authored the book “Machine Learning from Weak Supervision: An Empirical Risk Minimization Approach” (the MIT Press). On the other hand, he has served as an area chair 19 times, including ICLR 2021--2023, ICML 2019--2022, and NeurIPS 2019--2022. He also serves/has served as an action editor of TMLR and a guest editor of a special issue at MLJ. Moreover, he has served as a publication chair for ICML 2022, and has co-organized 11 workshops, 1 competition, and 3 tutorials.

\end{IEEEbiography}

\begin{IEEEbiography}[{\includegraphics[scale=0.12]{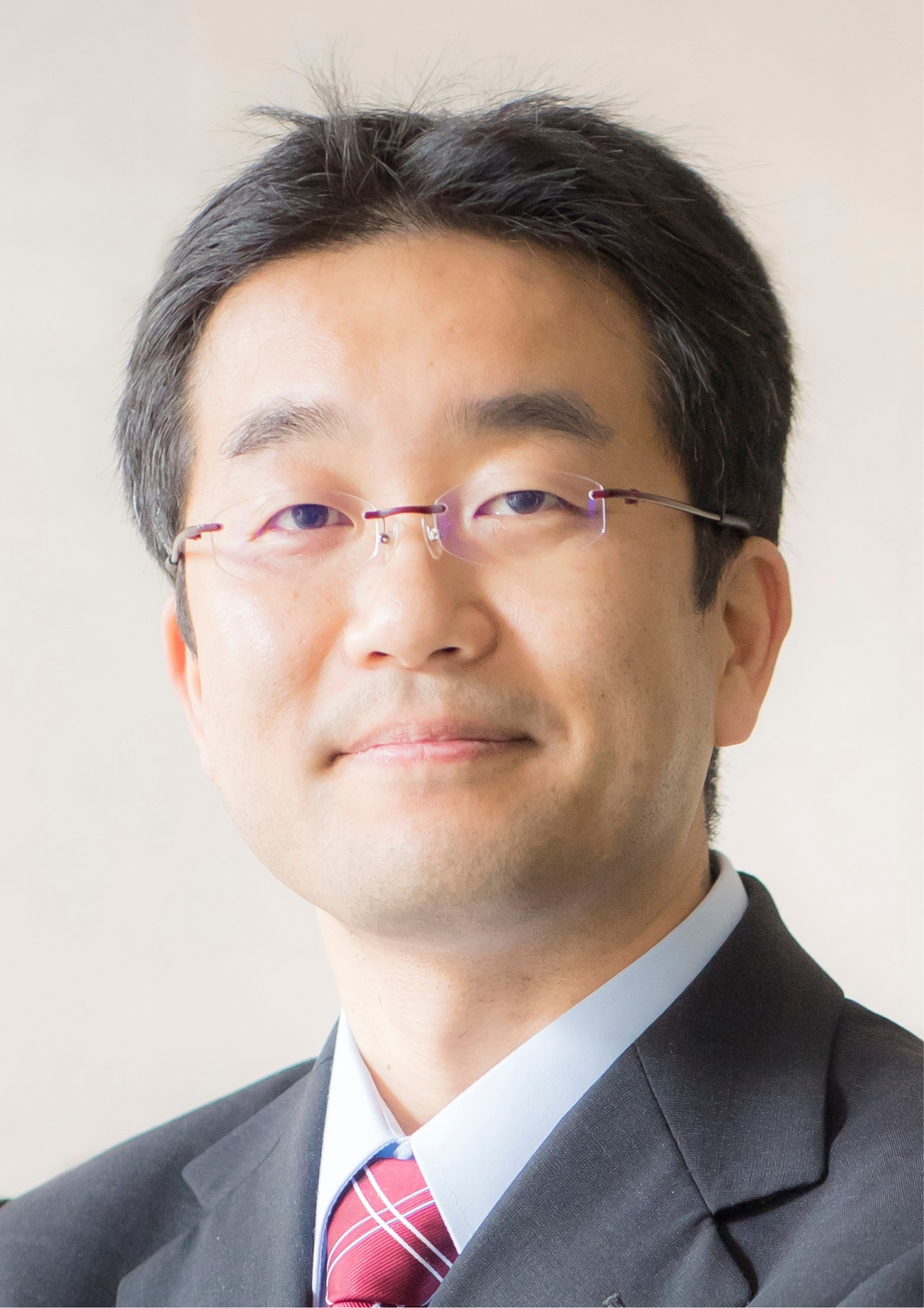}}]{Masashi Sugiyama}
received his Ph.D. in Computer Science from the Tokyo
Institute of Technology in 2001. He has been a professor at the
University of Tokyo since 2014, and also the director of the RIKEN
Center for Advanced Intelligence Project (AIP) since 2016. His
research interests include theories and algorithms of machine
learning. In 2022, he received the Award for Science and Technology
from the Japanese Minister of Education, Culture, Sports, Science and
Technology.
\end{IEEEbiography}







\begin{figure*}[h]
        \vspace{-2mm}
	\centering
	\includegraphics[scale=0.25]{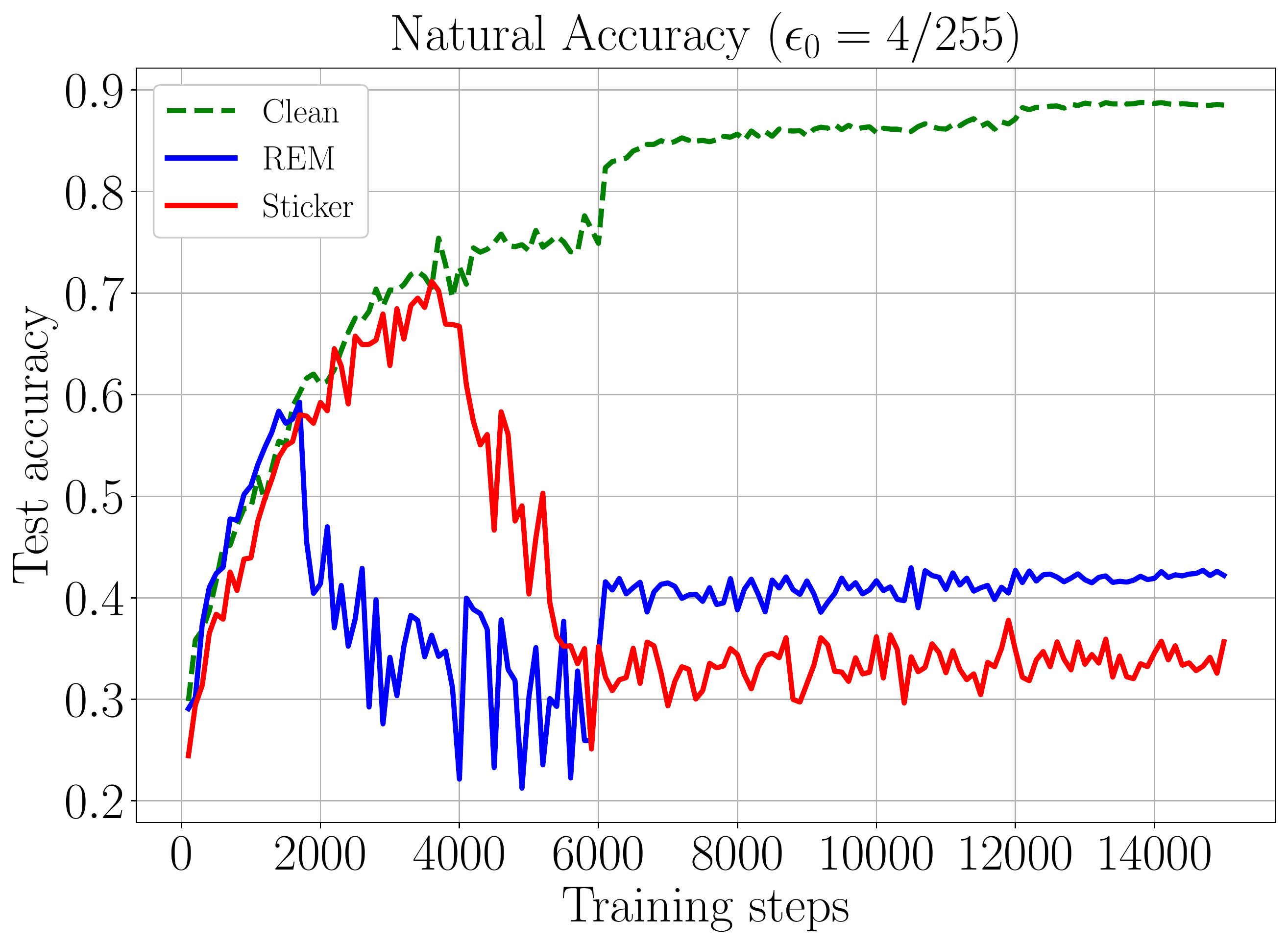}
	\hspace{15mm}
	\includegraphics[scale=0.25]{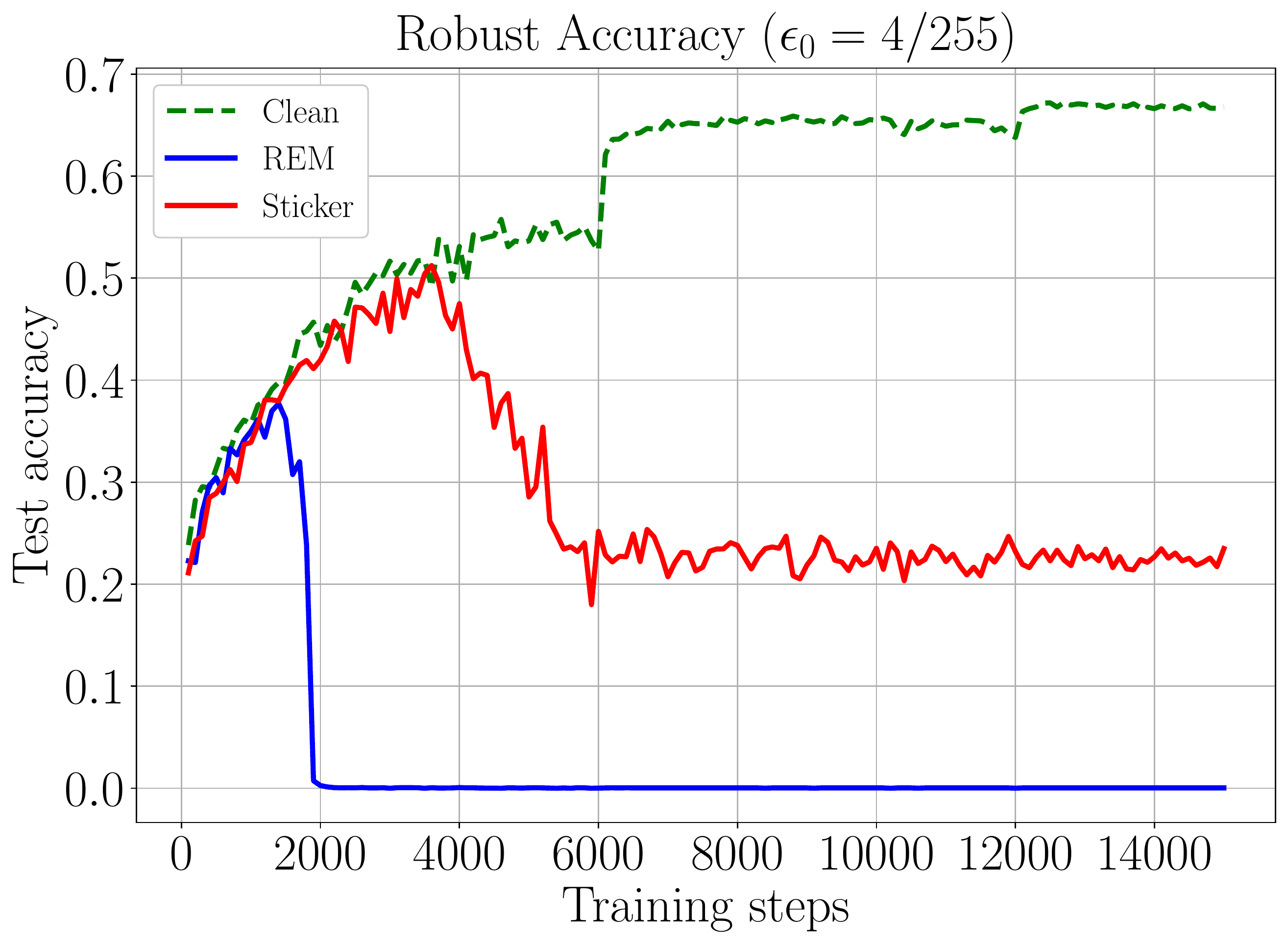} \\
	\includegraphics[scale=0.25]{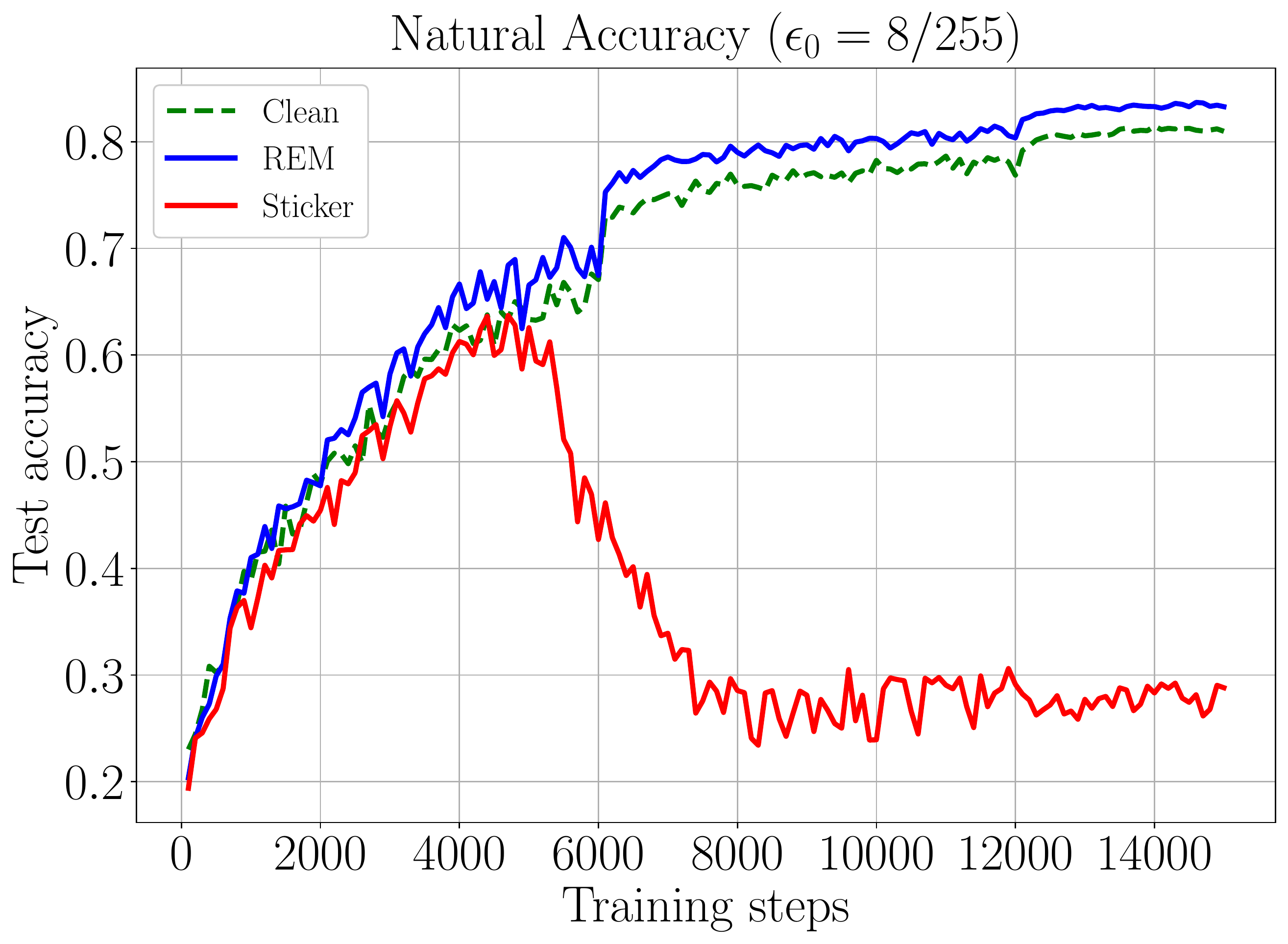}
	\hspace{15mm}
	\includegraphics[scale=0.25]{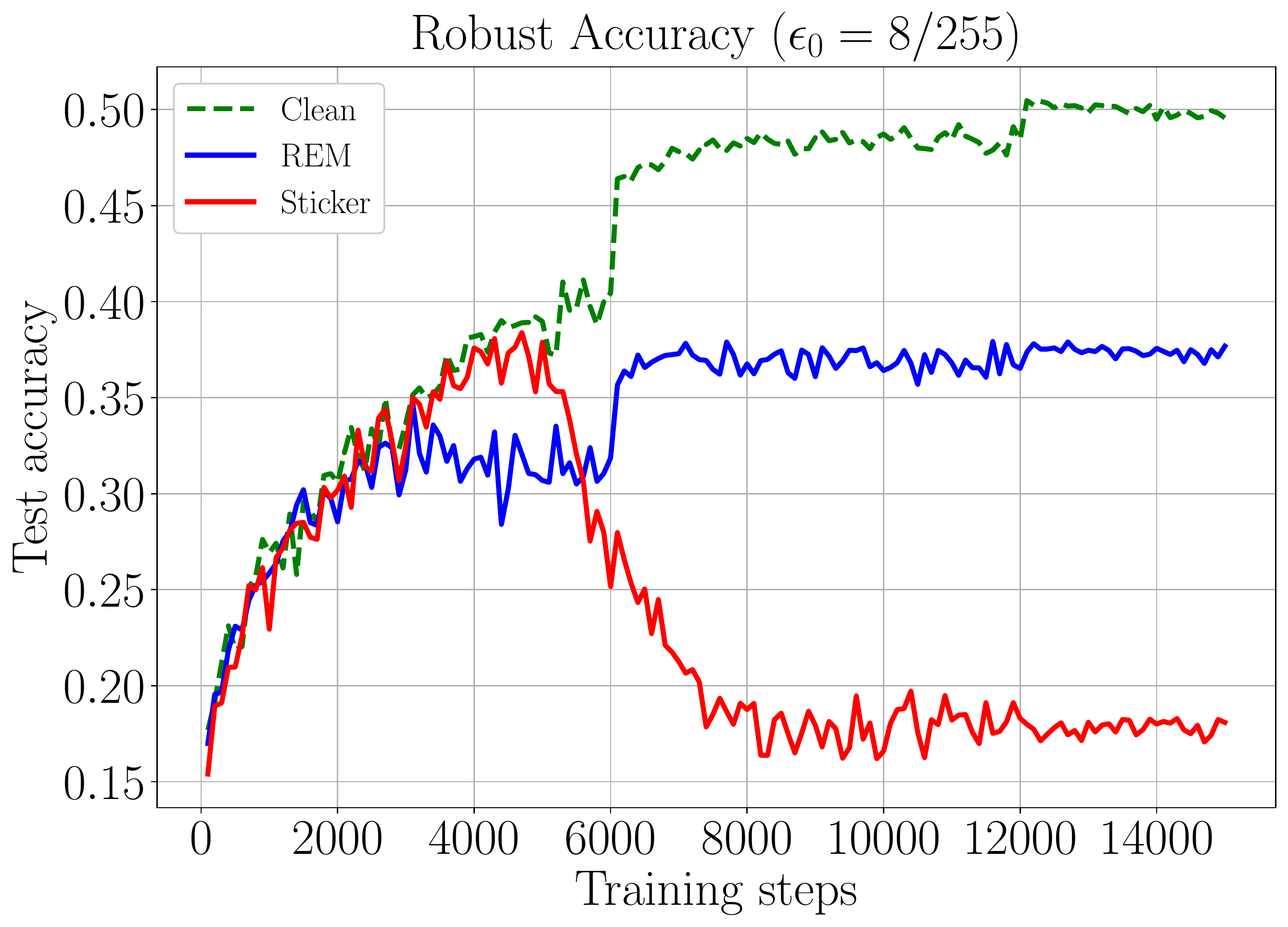} \\
	\includegraphics[scale=0.25]{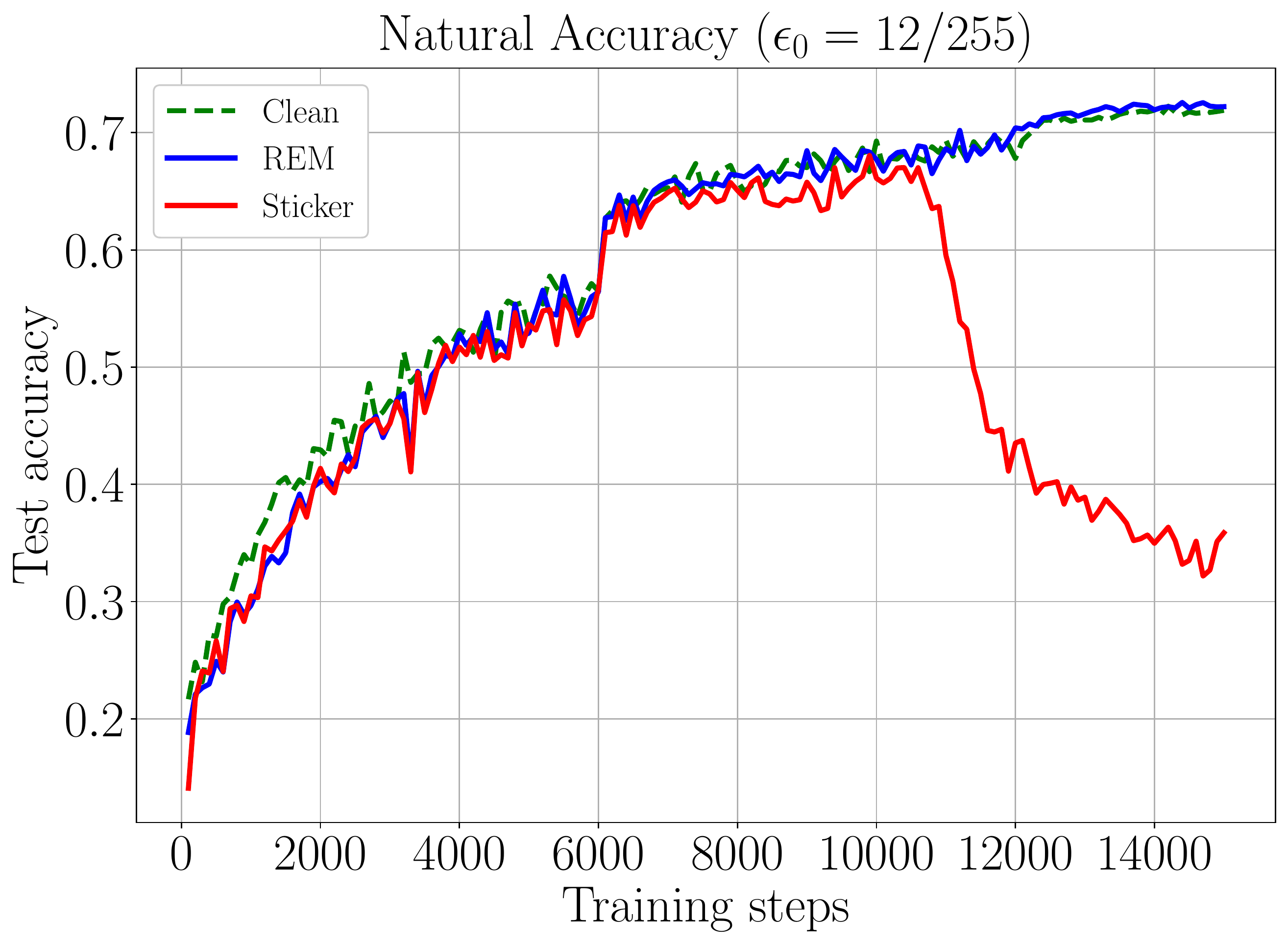}
	\hspace{15mm}
	\includegraphics[scale=0.25]{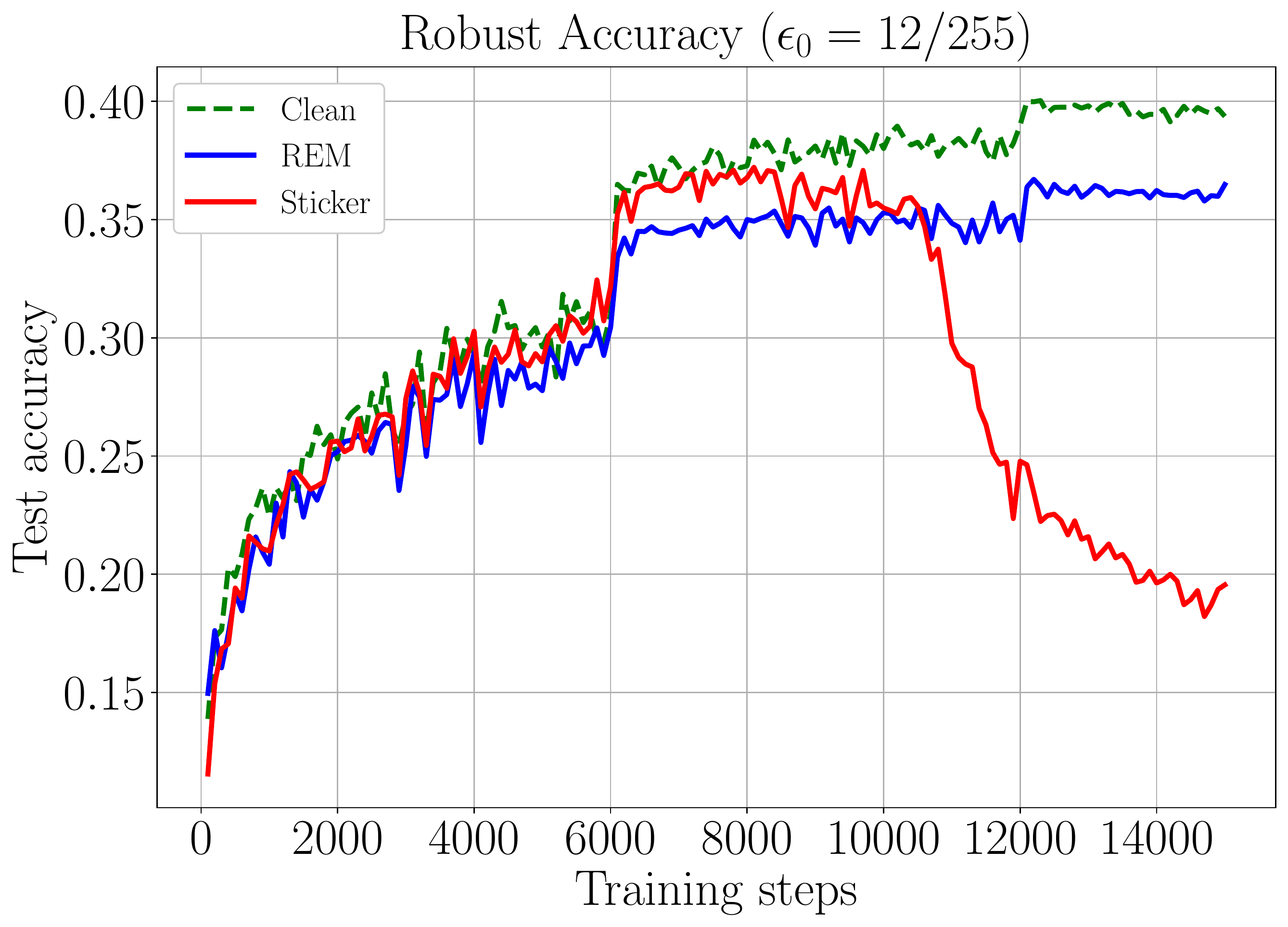} \\
	\includegraphics[scale=0.25]{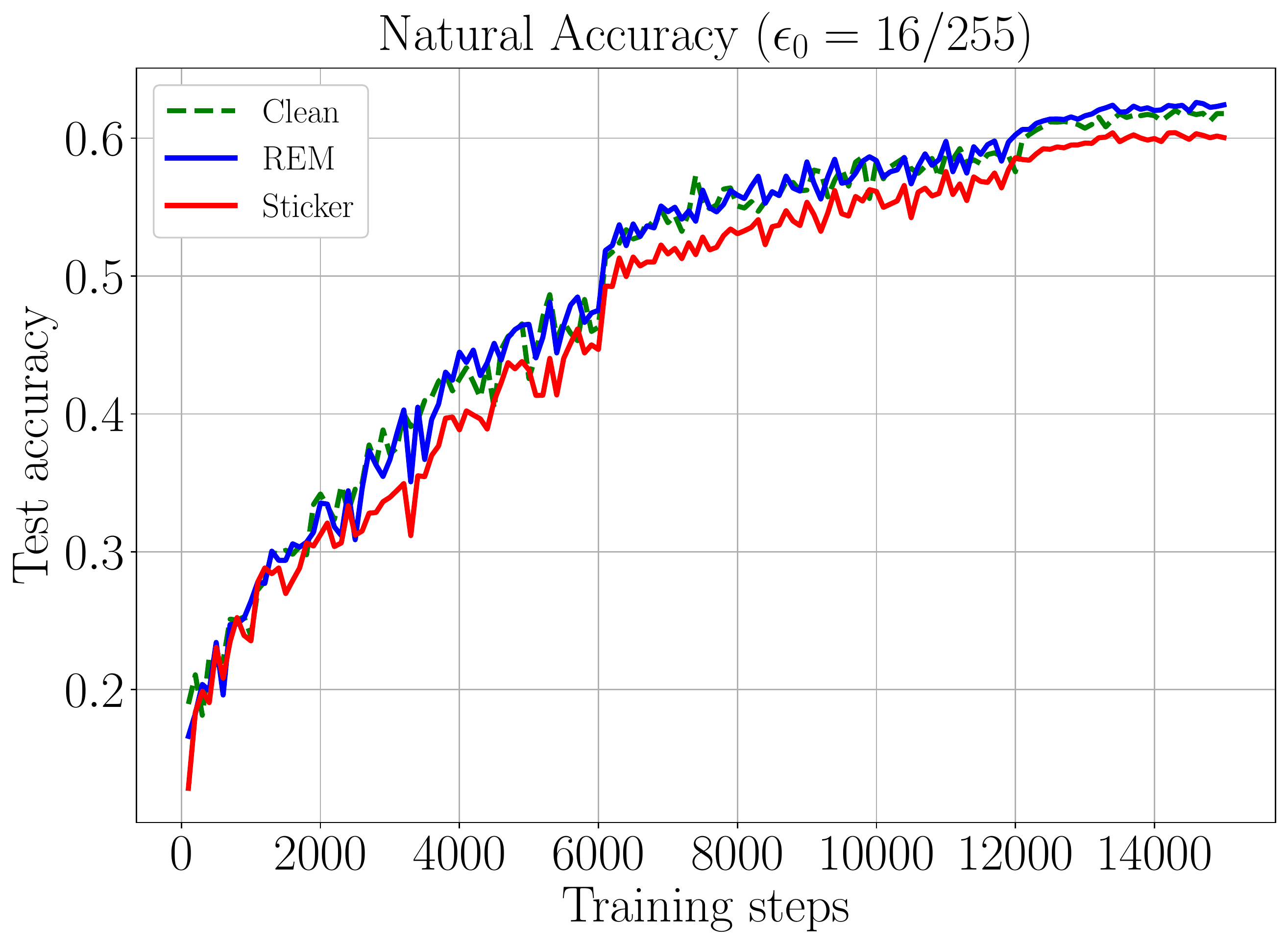}
	\hspace{15mm}
	\includegraphics[scale=0.25]{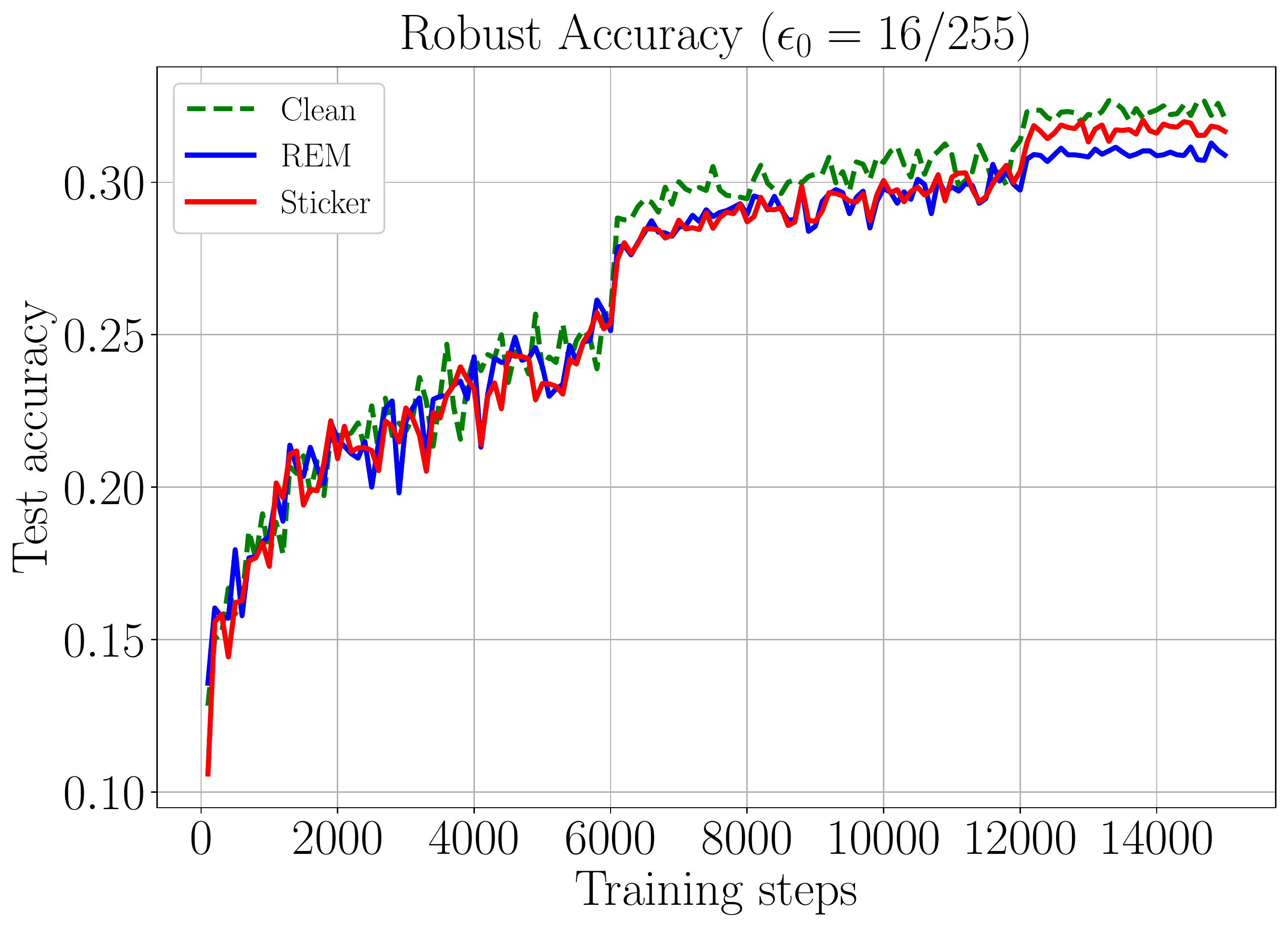} \\
    \vspace{-2mm}
	\caption{Test accuracy curves of robust learners $\A_{\epsilon_{0}}$ with $\epsilon_{0}$ in $\{4/255, 8/255, 12/255, 16/255\}$ on different sets. The green dashed lines are the oracle that uses $\epsballzero$-AT on the clean set $S$. The blue lines are $\epsballzero$-AT on the poisoned set $S'$ generated by REM, where $\epsball$ is $\ell_{\infty}$-norm ball with $\epsilon=8/255$. The red lines are $\epsballzero$-AT on the poisoned set $S'$ generated by Algorithm~\ref{alg:clean-label-untar} (sticker), where $\epsball'$ is $\ell_{0}$-norm ball, and the patch size takes $3\%$ of the whole pixels. }
	\label{fig:overfitting}
 \vspace{-2mm}
\end{figure*}

\newpage
\onecolumn
\appendix[Additional Experiment]
\noindent\textbf{Catastrophic overfitting gives a false sense of poisons.}
In Figure~\ref{fig:overfitting}, we report the full learning curves of the Table~\ref{exp:un-tar-attack_main-result} results on CIFAR-10.
We found there exists a phenomenon of catastrophic overfitting when the robust learner meets the poisoned data via both REM or our stickers, respectively (See $\epsilon_0 = \{4/255, 8/255, 12/255\}$). This overfitting can be easily combated by simply early stop the training process based on the validation set, which gives a false sense of poisoning effectiveness. 
Besides, we also found that when the robust learner employs very large $\epsilon_0$ (such as $16/255$), both REM and our stickers have little poisoning effect, which may need attackers to increase the poisoning radius $\epsilon$.

%

\end{document}